\documentclass[twoside,twocolumn,9pt]{article}
\usepackage{extsizes}
\usepackage[super,sort&compress,comma]{natbib} 
\usepackage[version=3]{mhchem}
\usepackage[left=1.5cm, right=1.5cm, top=1.785cm, bottom=2.0cm]{geometry}
\usepackage{balance}
\usepackage{widetext}
\usepackage{times,mathptmx}
\usepackage{sectsty}
\usepackage{graphicx} 
\usepackage{lastpage}
\usepackage[format=plain,justification=raggedright,singlelinecheck=false,font={stretch=1.125,small,sf},labelfont=bf,labelsep=space]{caption}
\usepackage{float}
\usepackage{fancyhdr}
\usepackage{fnpos}
\usepackage[english]{babel}
\usepackage{array}
\usepackage{droidsans}
\usepackage{charter}
\usepackage[T1]{fontenc}
\usepackage[usenames,dvipsnames]{xcolor}
\usepackage{setspace}
\usepackage[compact]{titlesec}

\definecolor{cream}{RGB}{222,217,201}

\begin{document}

\pagestyle{fancy}
\thispagestyle{plain}
\fancypagestyle{plain}{

\renewcommand{\headrulewidth}{0pt}
}

\makeFNbottom
\makeatletter
\renewcommand\LARGE{\@setfontsize\LARGE{15pt}{17}}
\renewcommand\Large{\@setfontsize\Large{12pt}{14}}
\renewcommand\large{\@setfontsize\large{10pt}{12}}
\renewcommand\footnotesize{\@setfontsize\footnotesize{7pt}{10}}
\makeatother

\renewcommand{\thefootnote}{\fnsymbol{footnote}}
\renewcommand\footnoterule{\vspace*{1pt}%
\color{cream}\hrule width 3.5in height 0.4pt \color{black}\vspace*{5pt}} 
\setcounter{secnumdepth}{5}

\makeatletter 
\renewcommand\@biblabel[1]{#1}            
\renewcommand\@makefntext[1]%
{\noindent\makebox[0pt][r]{\@thefnmark\,}#1}
\makeatother 
\renewcommand{\figurename}{\small{Fig.}~}
\sectionfont{\sffamily\Large}
\subsectionfont{\normalsize}
\subsubsectionfont{\bf}
\setstretch{1.125} 
\setlength{\skip\footins}{0.8cm}
\setlength{\footnotesep}{0.25cm}
\setlength{\jot}{10pt}
\titlespacing*{\section}{0pt}{4pt}{4pt}
\titlespacing*{\subsection}{0pt}{15pt}{1pt}

\fancyfoot{}
\fancyfoot[LO,RE]{\vspace{-7.1pt}\includegraphics[height=9pt]{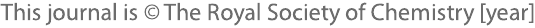}}
\fancyfoot[CO]{\vspace{-7.1pt}\hspace{13.2cm}\includegraphics{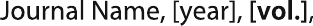}}
\fancyfoot[CE]{\vspace{-7.2pt}\hspace{-14.2cm}\includegraphics{head_foot/RF}}
\fancyfoot[RO]{\footnotesize{\sffamily{1--\pageref{LastPage} ~\textbar  \hspace{2pt}\thepage}}}
\fancyfoot[LE]{\footnotesize{\sffamily{\thepage~\textbar\hspace{3.45cm} 1--\pageref{LastPage}}}}
\fancyhead{}
\renewcommand{\headrulewidth}{0pt} 
\renewcommand{\footrulewidth}{0pt}
\setlength{\arrayrulewidth}{1pt}
\setlength{\columnsep}{6.5mm}
\setlength\bibsep{1pt}

\makeatletter 
\newlength{\figrulesep} 
\setlength{\figrulesep}{0.5\textfloatsep} 

\newcommand{\topfigrule}{\vspace*{-1pt}%
\noindent{\color{cream}\rule[-\figrulesep]{\columnwidth}{1.5pt}} }

\newcommand{\botfigrule}{\vspace*{-2pt}%
\noindent{\color{cream}\rule[\figrulesep]{\columnwidth}{1.5pt}} }

\newcommand{\dblfigrule}{\vspace*{-1pt}%
\noindent{\color{cream}\rule[-\figrulesep]{\textwidth}{1.5pt}} }

\makeatother

\twocolumn[
  \begin{@twocolumnfalse}
{\includegraphics[height=30pt]{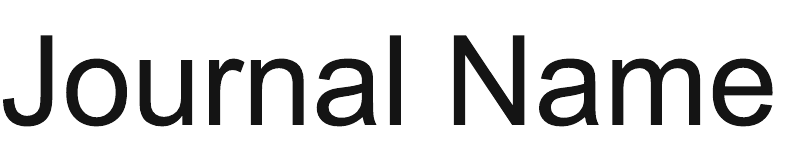}\hfill
 \raisebox{0pt}[0pt][0pt]{\includegraphics[height=55pt]{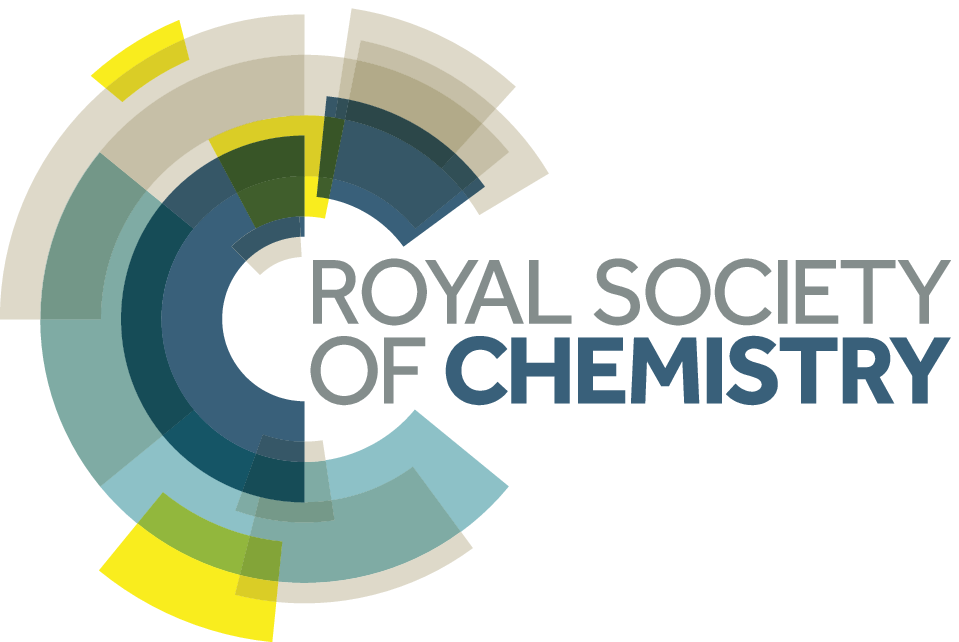}}
 \\[1ex]
 \includegraphics[width=18.5cm]{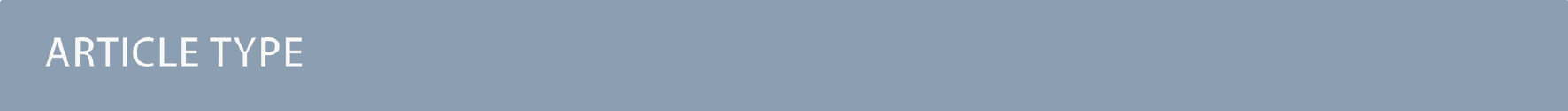}}\par
\vspace{1em}
\vspace{3cm}
\sffamily
\begin{tabular}{m{4.5cm} p{13.5cm} }

\includegraphics{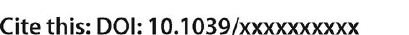} & \noindent\LARGE{\textbf{Geometry and mechanics of disclination lines in 3D nematic liquid crystals}} \\
\vspace{0.3cm} & \vspace{0.3cm} \\

 & \noindent\large{Cheng Long,\textit{$^{a}$} Xingzhou Tang,\textit{$^{a}$} Robin L. B. Selinger,\textit{$^{a}$} and Jonathan V. Selinger$^{\ast}$\textit{$^{a}$}} \\

\includegraphics{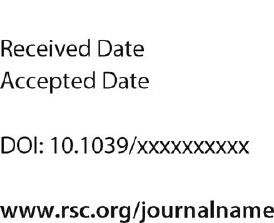} & \noindent\normalsize{In 3D nematic liquid crystals, disclination lines have a range of geometric structures.  Locally, they may resemble $+1/2$ or $-1/2$ defects in 2D nematic phases, or they may have 3D twist.  Here, we analyze the structure in terms of the director deformation modes around the disclination, as well as the nematic order tensor inside the disclination core.  Based on this analysis, we construct a vector to represent the orientation of the disclination, as well as tensors to represent higher-order structure.  We apply this method to simulations of a 3D disclination arch, and determine how the structure changes along the contour length.  We then use this geometric analysis to investigate three types of forces acting on a disclination:  Peach-Koehler forces due to external stress, interaction forces between disclination lines, and active forces.  These results apply to the motion of disclination lines in both conventional and active liquid crystals.} \\

\end{tabular}

 \end{@twocolumnfalse} \vspace{0.6cm}

  ]

\renewcommand*\rmdefault{bch}\normalfont\upshape
\rmfamily
\section*{}
\vspace{-1cm}


\footnotetext{\textit{$^{a}$~Department of Physics, Advanced Materials and Liquid Crystal Institute, Kent State University, Kent, OH 44242, USA; E-mail: jselinge@kent.edu}}





\section{Introduction}

Disclinations are the fundamental topological defects of nematic liquid crystals.  In conventional, passive liquid crystals, disclinations are important for the statistical mechanics of the isotropic-nematic transition, and for the coarsening dynamics of nematic order.  In active liquid crystals,\cite{Marchetti2013,Ramaswamy2017} disclinations are particularly important for the dynamic behavior because they are continually in motion, with disclination pairs nucleating and annihilating.  Hence, the physics of disclinations has been studied for many years,\cite{Friedel1969,Mermin1979,Kleman1983,Kleman1989,Chuang1991,Bowick1994,Terentjev1995,Kleman2008,Copar2011,Alexander2012,Copar2014} but especially in the recent context of active liquid crystals.\cite{Giomi2013,Pismen2013,Giomi2014,DeCamp2015,Vromans2016,Tang2017,Shankar2018,Kumar2018,Tang2019,Shankar2019}

Most research on active liquid crystals has concentrated on two-dimensional (2D) systems.  In 2D, disclinations have the structures shown in Fig.~1.  Here, the gray double-headed arrows represent the local nematic director field $\hat{\mathbf{n}}(\mathbf{r})$.  This director field is well-defined everywhere except at the singular points shown in black.  The point on the left is a disclination of topological charge $+1/2$, because the director rotates halfway around a circle in a positive sense as one moves around the blue loop.  Likewise, the point on the right is a disclination of topological charge $-1/2$, because the director rotates halfway around a circle in a negative sense around the blue loop.  Higher topological charges are mathematically possible, but they are rare because they have much higher energy.  The $+1/2$ disclination is shaped like a comet, with a characteristic orientation shown by the red arrow, which can be represented by a vector.\cite{Vromans2016}  The $-1/2$ disclination has three-fold symmetry, with a characteristic orientation shown by the red triad, which can be represented by a third-rank tensor.\cite{Tang2017}  In active liquid crystals, a $+1/2$ disclination induces a fluid flow pattern, which causes it to move along the direction of the red arrow (either forward or backward, depending on the type of activity).  This active motion has been investigated extensively through theory, simulations, and experiments.

\begin{figure}
\includegraphics[width=\columnwidth]{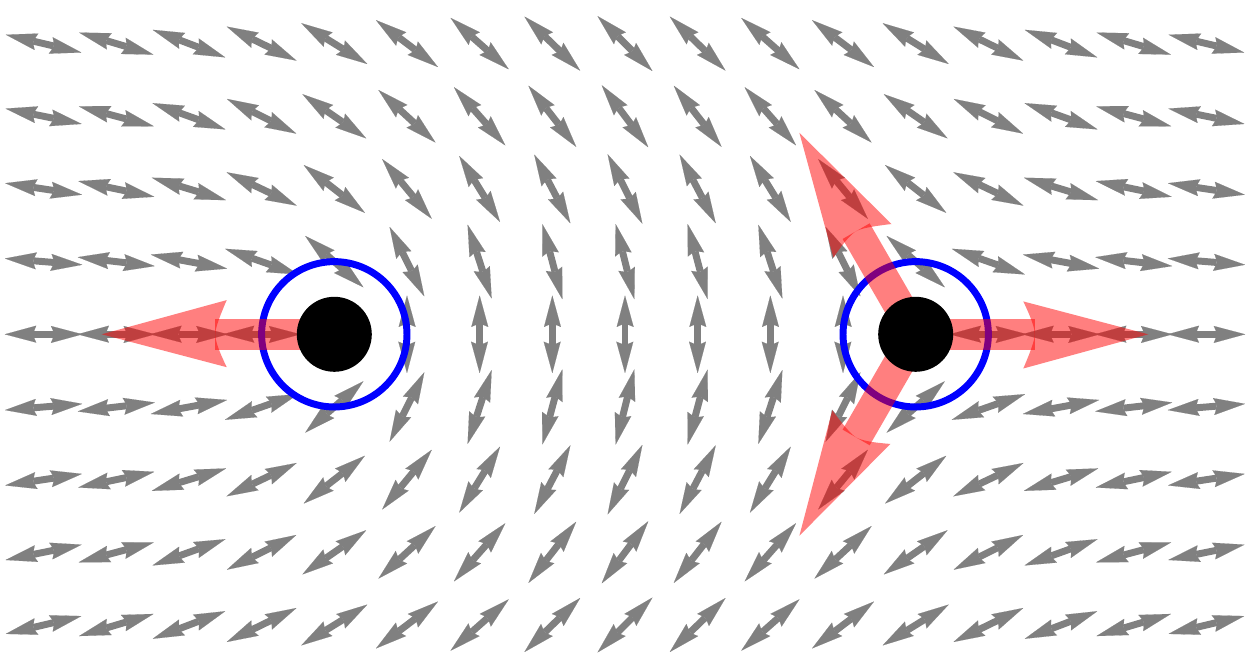}
\caption{Structure of a $+1/2$ disclination (left) and a $-1/2$ disclination (right) in a 2D nematic liquid crystal.}
\end{figure}

In the last two years, experiments have begun to study active nematic liquid crystals in three dimensions (3D),\cite{Duclos2020} and hence theories must investigate the physics of 3D disclinations.  The generalization from 2D to 3D is not trivial, because 3D disclination lines have very different topological properties than 2D disclination points.  When the nematic order is 2D, there are topologically distinct types of disclinations with any half-integer or integer charge.  However, when the nematic order is 3D, it is possible to continuously transform a $+1/2$ into a $-1/2$ disclination by twist of the director, and it is possible to continuously transform any integer disclination into a defect-free state by escape into the third dimension.  Hence, from a topological perspective, there is only one type of 3D disclination line, and active nematic liquid crystals must be understood in terms of this one type of disclination line.

Two recent articles have made important contributions to the theory of disclination lines in 3D active nematic liquid crystals.  Duclos \emph{et al}.\cite{Duclos2020} provide a combined experimental and theoretical study.  On the theoretical side, they develop a geometric method to characterize disclinations, and show that the local structure of a disclination can be described in terms of certain vectors.  From the relationship among these vectors, one can distinguish whether the local director field has a $+1/2$ wedge (planar) structure, a twisted 3D structure, a $-1/2$ wedge structure, or something intermediate between these cases.  Moreover, by considering how these vectors change around the entire length of a closed disclination loop, one can characterize the topological properties of the whole loop.  In related work, based on a large-scale computational study,\cite{Copar2019} Binysh \emph{et al}.\cite{Binysh2020} develop a theory for the dynamics of disclinations in 3D active nematic systems.  They put the director field into the Stokes equation for fluid flow driven by the active force, and calculate the self-propelled velocity of a local segment of a disclination line.  They then consider the dynamic properties of a closed disclination loop, and show that the activity might drive the loop to extend, or contract, or buckle into a nonplanar 3D shape.

The purpose of this paper is to investigate the geometric structure of a disclination line, and the forces acting on a disclination line, in ways that complement those previous articles.  We begin by characterizing the orientational properties of a 3D disclination.  {\v{C}}opar \emph{et al.}\cite{Copar2011} have visualized disclinations as ribbons with orientational properties; here we construct mathematical objects to represent such properties.  In particular, we generalize the previous construction of a vector orientation for a planar $+1/2$ disclination, or a third-rank tensor orientation for a planar $-1/2$ disclination.  In Sec.~2 we address that problem by calculating the director deformations \emph{around} a disclination line, and in Sec.~3 we consider the same problem from the perspective of the nematic order tensor $Q_{ij}$ \emph{inside} the disclination core.  In both cases, we find that the orientational properties decompose into a vector, a second-rank tensor, and a third-rank tensor.  In the limiting case of a planar structure, the vector becomes the orientation for a $+1/2$ disclination, and the third-rank tensor becomes the orientation for a $-1/2$ disclination.  The second-rank tensor occurs in 3D but not in 2D, and is associated with a twisted disclination.  We use this geometric construction to analyze simulations in Sec.~4.

As a further step, we use the geometric construction to determine the forces acting on a disclination line.  We begin by considering an analogy with dislocation lines in crystalline solids.  In solids, the Peach-Koehler force is the force on a dislocation line due to an applied shear stress.  The concept of a Peach-Koehler force was applied to nematic liquid crystals by Kl\'{e}man.\cite{Kleman1983}  In Sec.~5, we analyze examples of a Peach-Koehler force in a nematic liquid crystal under an applied rotational stress.  In solids, the Peach-Koehler force can be used to determine the force of one dislocation line on another.  In Sec.~6, we perform the analogous calculation in a nematic liquid crystal, to find the the force of one disclination line on another.  Finally, in Sec.~7, we consider the active force on a disclination line in a 3D active liquid crystal.  Our result is equivalent to Binysh \emph{et al}.,\cite{Binysh2020} and we discuss it in terms of the geometric construction for orientational properties of a disclination.

\section{Director deformations around disclination}

In this section, we regard a disclination line as a region of a liquid crystal with high director deformations.  With that point of view, we want to characterize the disclination line by characterizing the director deformations.  Hence, we describe the director field around a disclination using a parameterization equivalent to Binysh \emph{et al}.,\cite{Binysh2020} and then calculate the relevant derivatives.

For this parameterization, we assume the disclination line has a local tangent vector $\hat{\mathbf{t}}$.  Without loss of generality, we choose the $z$-axis to lie along $\hat{\mathbf{t}}$.  As the position $\mathbf{r}$ moves in a loop about $\hat{\mathbf{t}}$, the director field $\hat{\mathbf{n}}(\mathbf{r})$ rotates through a half-circle.  Let us consider the minimal distortion state, i.e.\ the ground state for a liquid crystal with equal Frank constants.  In this state, the half-circle lies in a plane, which is characterized by two orthonormal vectors $\hat{\mathbf{m}}$ and $\hat{\mathbf{m}}'$.  We choose $\hat{\mathbf{m}}$ to be the direction where the plane crosses the equator, perpendicular to $\hat{\mathbf{t}}$, and $\hat{\mathbf{m}}'$ to be the direction where the plane is farthest from the equator.  Hence, the director field can be written as
\begin{equation}
\hat{\mathbf{n}}=\hat{\mathbf{m}}\cos\frac{\phi-\phi_0}{2}+\hat{\mathbf{m}}'\sin\frac{\phi-\phi_0}{2},
\label{director}
\end{equation}
where $\phi$ is the azimuthal angle about the defect line, and $\phi_0$ is the spatial direction in which $\hat{\mathbf{n}}=\hat{\mathbf{m}}$.  This director field is illustrated in Fig.~2.

\begin{figure}
\begin{center}
\includegraphics[width=.6\columnwidth]{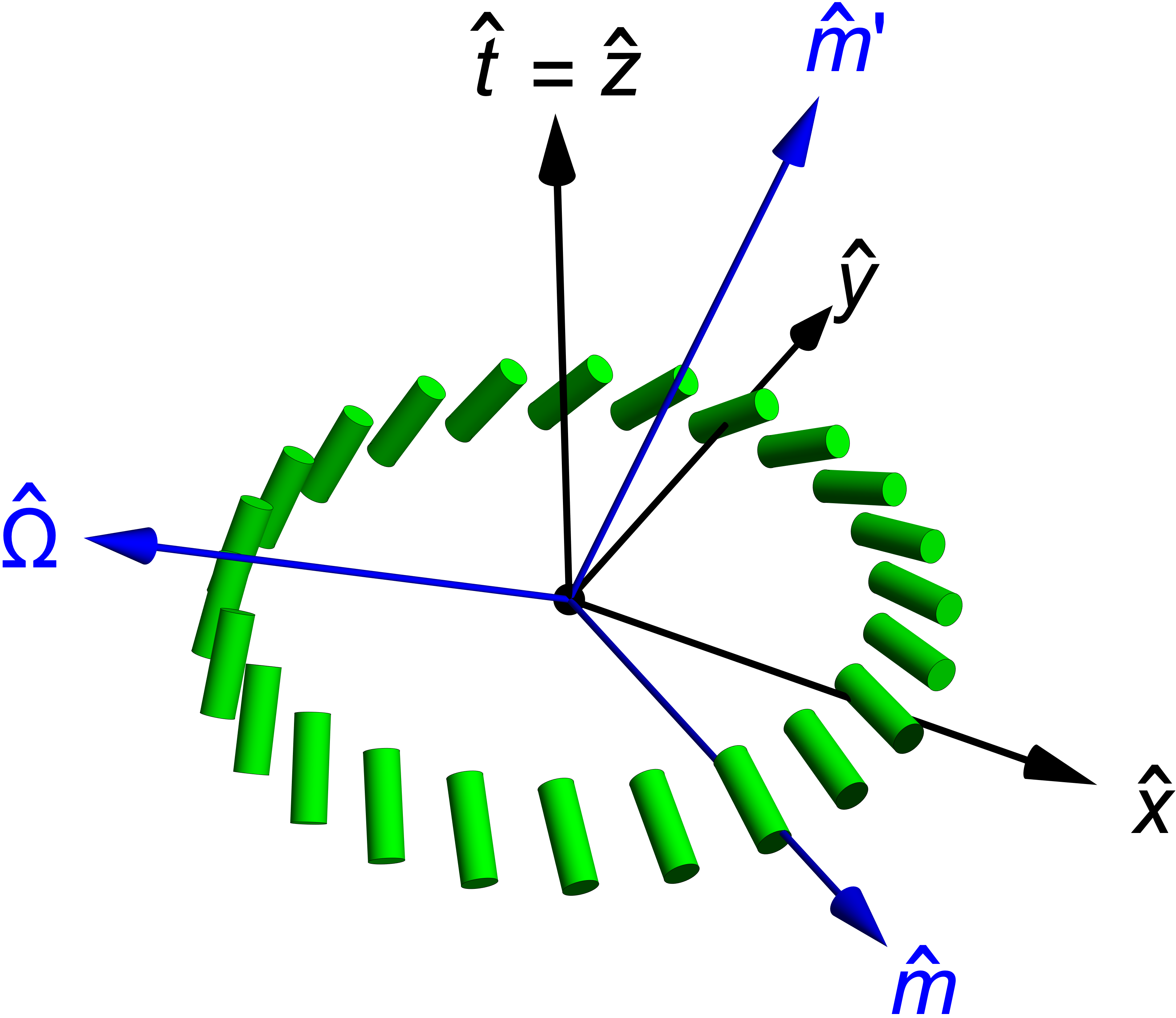}
\end{center}
\caption{Geometry of a disclination in a 3D nematic liquid crystal, with the angles $\alpha=-\pi/6$, $\beta=\pi/3$, and $\phi_0 = 0$.  The director $\hat{\mathbf{n}}$ lies in the $(\hat{\mathbf{m}},\hat{\mathbf{m}}')$ plane, with $\hat{\mathbf{n}}=\hat{\mathbf{m}}$ for $\phi=\phi_0$, and $\hat{\mathbf{n}}=\hat{\mathbf{m}}'$ for $\phi=\phi_0 +\pi$.}
\end{figure}

From Eq.~(\ref{director}), we see that the director field $\hat{\mathbf{n}}$ rotates about the vector $\hat{\pmb{\Omega}}=\hat{\mathbf{m}}\times\hat{\mathbf{m}}'$.  This rotation vector $\hat{\pmb{\Omega}}$ was first defined by Friedel and de Gennes.\cite{Friedel1969}  In general, $\hat{\pmb{\Omega}}$ may be at any angle $\beta$ with respect to $\hat{\mathbf{t}}$.  Hence, the orthonormal triad $(\hat{\mathbf{m}},\hat{\mathbf{m}}',\hat{\pmb{\Omega}})$ can be written explicitly as
\begin{align}
\label{orthonormaltriad}
&\hat{\mathbf{m}}=(\cos\alpha,\sin\alpha,0),\\
&\hat{\pmb{\Omega}}=\hat{\mathbf{t}}\cos\beta+(\hat{\mathbf{m}}\times\hat{\mathbf{t}})\sin\beta
=(\sin\beta\sin\alpha,-\sin\beta\cos\alpha,\cos\beta),\nonumber\\
&\hat{\mathbf{m}}'=\hat{\pmb{\Omega}}\times\hat{\mathbf{m}}
=(-\cos\beta\sin\alpha,\cos\beta\cos\alpha,\sin\beta).\nonumber
\end{align}
The director field with $\beta=0$ ($\hat{\pmb{\Omega}}$ parallel to $\hat{\mathbf{t}}$) is equivalent to the $+1/2$ planar disclination shown on the left of Fig.~1.  Likewise, the director field with $\beta=\pi$ ($\hat{\pmb{\Omega}}$ antiparallel to $\hat{\mathbf{t}}$) is equivalent to the $-1/2$ planar disclination shown on the right of Fig.~1.  Those two planar cases are both called ``wedge'' disclinations.  By comparison, an intermediate angle $\beta$ gives a 3D nonplanar disclination, which continuously interpolates between the $+1/2$ and $-1/2$ limits.  In particular, the case of $\beta=\pi/2$ ($\hat{\pmb{\Omega}}$ perpendicular to $\hat{\mathbf{t}}$) is called a ``twist'' disclination.

We now analyze the gradients of this director field.  For this analysis, we use a mathematical formalism which was recently developed by Machon and Alexander\cite{Machon2016} and applied to elasticity theory by Selinger.\cite{Selinger2018}  The formalism decomposes the director gradient tensor into four modes,
\begin{equation}
\partial_i n_j =-n_i B_j +\frac{1}{2}T\epsilon_{ijk}n_k +\frac{1}{2}S(\delta_{ij}-n_i n_j )+\Delta_{ij}.
\end{equation}
The first three modes are the well-known bend, twist, and splay deformations,
\begin{equation}
\mathbf{B}=\hat{\mathbf{n}}\times(\mathbf{\nabla}\times\hat{\mathbf{n}}),\qquad
T=\hat{\mathbf{n}}\cdot(\mathbf{\nabla}\times\hat{\mathbf{n}}),\qquad
S=\mathbf{\nabla}\cdot\hat{\mathbf{n}}.
\label{btsdefinitions}
\end{equation}
The fourth mode $\Delta_{ij}$ is a tensor deformation mode, which is less well-known but is related to saddle-splay.  It has been called ``anisotropic orthogonal gradients of $\hat{\mathbf{n}}$''\cite{Machon2016} or ``biaxial splay,''\cite{Selinger2018} and it can be written mathematically as
\begin{align}
\Delta_{ij} &= \frac{1}{2} [\partial_i n_j + \partial_j n_i + n_i B_j + n_j B_i - S (\delta_{ij} - n_i n_j)]\\
&= \frac{1}{2} [\partial_i n_j + \partial_j n_i - n_i n_k \partial_k n_j - n_j n_k \partial_k n_i
- \delta_{ij} \partial_k n_k + n_i n_j \partial_k n_k ].\nonumber
\end{align}

For the simplest deformation mode, let us begin with twist.  From the definition in Eq.~(\ref{btsdefinitions}), we see that the twist pseudoscalar $T$ is even in the director $\hat{\mathbf{n}}$.  For that reason, $T$ is uniquely defined, in spite of the fact that the sign of $\hat{\mathbf{n}}$ is not uniquely defined; $+\hat{\mathbf{n}}$ and $-\hat{\mathbf{n}}$ both represent the same physical state.  From the director field of Eqs.~(\ref{director}--\ref{orthonormaltriad}), we can calculate explicitly
\begin{equation}
T=\frac{\sin\beta\cos(\phi-\alpha)}{2\rho},
\end{equation}
where $\rho$ is the radius outward from the disclination line in cylindrical coordinates.  Clearly this twist depends on the azimuthal angle $\phi$; there are regions of positive and negative twist.  For that reason, we integrate around a loop at fixed $\rho$ to obtain the average centered on the disclination,
\begin{equation}
\overline{T}=\frac{1}{2\pi}\int_{-\pi}^{\pi}T d\phi=0.
\end{equation}
This average is zero for all $\beta$, even for the twist disclination with $\beta=\pi/2$, because the regions of positive and negative twist exactly cancel each other.  Hence, we will not consider the twist mode further.

Next, consider the bend deformation mode.  Like the twist, the bend vector $\mathbf{B}$ is even in $\hat{\mathbf{n}}$, and hence $\mathbf{B}$ is uniquely defined under exchange of $\pm\hat{\mathbf{n}}$.  From the director field of Eqs.~(\ref{director}--\ref{orthonormaltriad}), we calculate $\mathbf{B}(\rho,\phi)$, and then average around a loop at fixed $\rho$.  The result is
\begin{equation}
\overline{\mathbf{B}}=-\frac{1+\cos\beta}{8\rho}(\hat{\mathbf{m}}\cos\gamma+\hat{\mathbf{m}}'\sin\gamma),
\end{equation}
where $\gamma=\alpha-\phi_0$.  From these results, we can see that the average bend $\overline{\mathbf{B}}$ lies in the $(\hat{\mathbf{m}},\hat{\mathbf{m}}')$ plane, perpendicular to $\hat{\pmb{\Omega}}$.  It is proportional to $1+\cos\beta$, and hence it is largest for a wedge $+1/2$ disclination with $\beta=0$, it has an intermediate value for a twist disclination with $\beta=\pi/2$, and it vanishes for a wedge $-1/2$ disclination $\beta=\pi$.  It scales inversely with the radius $\rho$ outward from the disclination line.

For the third deformation mode, we have splay.  This case has an extra subtlety, because the splay scalar $S$ is odd in the director $\hat{\mathbf{n}}$.  For that reason, the sign of $S$ is not uniquely defined, and averaging $S$ around a disclination is not meaningful.  However, the splay vector $\mathbf{S}=S\hat{\mathbf{n}}$ is even in $\hat{\mathbf{n}}$, so it is uniquely defined.  Hence, we calculate $S\hat{\mathbf{n}}$  from the director field of Eqs.~(\ref{director}--\ref{orthonormaltriad}), and average it around a loop at fixed $\rho$, to obtain
\begin{equation}
\overline{S\hat{\mathbf{n}}}=\frac{1+\cos\beta}{8\rho}(\hat{\mathbf{m}}\cos\gamma+\hat{\mathbf{m}}'\sin\gamma).
\end{equation}
This result for $\overline{S\hat{\mathbf{n}}}$ is exactly the negative of the result for $\overline{\mathbf{B}}$, and thus it carries the same information.  From the results for bend and splay, we see that it is useful to define the vector
\begin{equation}
\mathbf{p}=\frac{1+\cos\beta}{2}(\hat{\mathbf{m}}\cos\gamma+\hat{\mathbf{m}}'\sin\gamma),
\label{pdefinition}
\end{equation}
so that $\overline{S\hat{\mathbf{n}}}=-\overline{\mathbf{B}}=\mathbf{p}/(4\rho)$.

Finally, we have the fourth deformation mode $\Delta_{ij}$.  Like the splay scalar $S$, the second-rank tensor $\Delta_{ij}$ is odd in the director $\hat{\mathbf{n}}$, and hence its sign is not uniquely defined.  However, the third-rank tensor $\Delta_{ij}n_k$ is even in $\hat{\mathbf{n}}$, so it is uniquely defined.  We calculate this third-rank tensor for the director field of Eqs.~(\ref{director}--\ref{orthonormaltriad}) and average it around a loop.  The resulting expression for $\overline{\Delta_{ij}n_k}$ is fairly long.  To interpret it, we break the expression into three components
\begin{equation}
\overline{\Delta_{ij}n_k}=\overline{\Delta_{ij}n_k}^{(1)}+\overline{\Delta_{ij}n_k}^{(2)}+\overline{\Delta_{ij}n_k}^{(3)},
\end{equation}
based on how they transform under rotations about $\hat{\pmb{\Omega}}$.  To be specific, we define the rotation matrix $R_{ij}(\phi)=\cos\phi (m_i m_j + m'_i m'_j)+\sin\phi (m'_i m_j - m_i m'_j) + \Omega_i \Omega_j$, and then calculate the components
\begin{equation}
\overline{\Delta_{ij}n_k}^{(l)}=\frac{1}{\pi}\int_{-\pi}^{\pi}d\phi\cos(l\phi)R_{ii'}(\phi)R_{jj'}(\phi)R_{kk'}(\phi)\overline{\Delta_{i'j'}n_{k'}},
\end{equation}
for $l=1$, $2$, and $3$.

The first component transforms as a vector under rotations about $\hat{\pmb{\Omega}}$.  It can be expressed as
\begin{equation}
\overline{\Delta_{ij}n_k}^{(1)}=\frac{(3\delta_{ij}-7\Omega_i\Omega_j)p_k-(\delta_{ik}-\Omega_i\Omega_k)p_j-(\delta_{jk}-\Omega_j\Omega_k)p_i}{32\rho}
\end{equation}
This component is proportional to $(1+\cos\beta)/2$; that factor is included in the definition of $\mathbf{p}$ in Eq.~(\ref{pdefinition}).  Like the bend and splay, this component is largest for a wedge $+1/2$ disclination, and it vanishes for a wedge $-1/2$ disclination.  It has a characteristic orientation given by the vector $\mathbf{p}$ in the plane perpendicular to $\hat{\pmb{\Omega}}$.

The second component transforms as a second-rank tensor under rotations about $\hat{\pmb{\Omega}}$.  It can be written as
\begin{equation}
\overline{\Delta_{ij}n_k}^{(2)}=\frac{\Omega_i D_{jk}+\Omega_j D_{ik}}{16\rho},
\end{equation}
where
\begin{equation}
D_{ij}=\sin\beta\left[(m_i m_j -m'_i m'_j)\sin\gamma-(m_i m'_j +m'_i m_j)\cos\gamma\right].
\label{Ddefinition}
\end{equation}
This component is proportional to $\sin\beta$.  It is largest for a twist disclination with $\beta=\pi/2$, and it vanishes for wedge $+1/2$ and $-1/2$ disclinations with $\beta=0$ and $\pi$.  In this component, $D_{ij}$ is a symmetric second-rank tensor in the plane perpendicular to $\hat{\pmb{\Omega}}$.  The eigenvalues of this tensor are $\pm\sin\beta$, and the corresponding eigenvectors identify the characteristic directions associated with a twist disclination.  (There is also a trivial eigenvalue of $0$, corresponding to the eigenvector $\hat{\pmb{\Omega}}$.)

The third component transforms as a third-rank tensor under rotations about $\hat{\pmb{\Omega}}$,
\begin{equation}
\overline{\Delta_{ij}n_k}^{(3)}=\frac{1}{32\rho}T_{ijk},
\end{equation}
where
\begin{align}
T_{ijk}=&\frac{1-\cos\beta}{2}\bigl[
(m_i m_j m_k - m_i m'_j m'_k - m'_i m_j m'_k - m'_i m'_j m_k)\cos\gamma\nonumber\\
&+(m_i m_j m'_k + m_i m'_j m_k + m'_i m_j m_k - m'_i m'_j m'_k)\sin\gamma\bigr].
\label{Tdefinition}
\end{align}
This component is proportional to $(1-\cos\beta)/2$.  It is largest for a wedge $-1/2$ disclination with $\beta=\pi$, and it vanishes for a wedge $+1/2$ disclination with $\beta=0$.  In this component, $T_{ijk}$ is a completely symmetric third-rank tensor in the plane perpendicular to $\hat{\pmb{\Omega}}$.  Hence, it identifies the characteristic directions associated with the wedge $-1/2$ disclination.

We will show visualizations of these geometric features for some sample disclination lines in Sec.~4.  Before that, it is useful to compare the 3D director gradients calculated here with the previous theory of defect orientation in 2D nematic liquid crystals.\cite{Vromans2016,Tang2017}  

In the limit of $\beta=0$, the director lies in the $(x,y)$ plane with $\hat{\pmb{\Omega}}=\hat{\mathbf{t}}$, and the 3D disclination becomes a 2D disclination with topological charge $+1/2$.  For this limiting case, the tensors $D_{ij}$ and $T_{ijk}$ both vanish, and the director gradients are characterized by the vector $\mathbf{p}$, with unit magnitude.  This vector then becomes identical to the 2D defect orientation vector, which is indicated by the red arrow on the left side of Fig.~1.

By comparison, in the limit of $\beta=\pi$, the director lies in the $(x,y)$ plane with $\hat{\pmb{\Omega}}=-\hat{\mathbf{t}}$, and the 3D disclination becomes a 2D disclination with topological charge $-1/2$.  In that case, the vector $\mathbf{p}$ and tensor $D_{ij}$ both vanish, and the director gradients are characterized by $T_{ijk}$, which becomes identical to the 2D defect orientation tensor for a $-1/2$ defect.  As discussed previously,\cite{Tang2017} that completely symmetric, third-rank tensor is associated with a triad of three orientations in the plane, $2\pi/3$ from each other, as indicated by the red triad on the right side of Fig.~1.

From that comparison, we can see that the 3D theory provides a generalization of the previous 2D theory of defect orientation.  It shows that the vector $\mathbf{p}$ and tensor $T_{ijk}$ smoothly interpolate between the $+1/2$ and $-1/2$ defects in 2D, and the tensor $D_{ij}$ gives an extra two-fold symmetric component of the orientation, which occurs in 3D but not in 2D. 

\section{Nematic order inside disclination core}

For an alternative perspective on disclination lines, we consider the nematic order tensor that is inside the core of a disclination line.

In general, nematic order is represented by a tensor field $Q_{ij}(\mathbf{r})$.  In the bulk, away from disclination lines, this tensor field is related to the director field by $Q_{ij}=s_\mathrm{bulk}(\frac{3}{2}n_i n_j -\frac{1}{2}\delta_{ij})$, with $s_\mathrm{bulk}>0$.  In those regions, it is a uniaxial tensor with eigenvalues $+s_\mathrm{bulk}$, $-s_\mathrm{bulk}/2$, and $-s_\mathrm{bulk}/2$, which are determined by minimizing the bulk free energy.  However, the form of this tensor changes inside of a disclination core.  The eigenvalues may differ from the bulk eigenvalues, and the tensor does not need to be uniaxial.  Indeed, classic theoretical work by Schopohl and Sluckin\cite{Schopohl1987} shows that this tensor is biaxial in most of the disclination core, and it is uniaxial with a negative order parameter in the exact center.

We would like to propose a model for the full tensor structure of $Q_{ij}$ inside the core, and use this model to describe orientational features of the disclination line.  For this construction, we begin by considering the $Q_{ij}$ tensor \emph{outside} the core, i.e. for $\rho>\xi$, where $\xi$ is the core radius.  The director field around a disclination is given by Eq.~(\ref{director}).  From that director field, the $Q_{ij}$ tensor becomes
\begin{align}
Q_{ij}=s_\mathrm{bulk}\biggl[&\frac{1}{4}\delta_{ij}-\frac{3}{4}\Omega_i \Omega_j +\frac{3}{4}(m_i m_j -m'_i m'_j)\cos(\phi-\phi_0)\nonumber\\
&+\frac{3}{4}(m_i m'_j +m'_i m_j)\sin(\phi-\phi_0)\biggr], \textrm{ for } \rho>\xi.
\label{Qoutsidepreliminary}
\end{align}
To simplify that expression, we define the position vector in cylindrical coordinates, $\mathbf{r} = (x,y,z) = (\rho\cos\phi,\rho\sin\phi,z)$.  We further define $\hat{\mathbf{v}}=(\cos\phi_0,\sin\phi_0,0)$ as the direction outward from the disclination core such that $\hat{\mathbf{n}}=\hat{\mathbf{m}}$ is in the equatorial plane, perpendicular to $\hat{\mathbf{t}}$, and $\hat{\mathbf{v}}'=\hat{\mathbf{t}}\times\hat{\mathbf{v}}=(-\sin\phi_0,\cos\phi_0,0)$ as an orthogonal vector outward from the disclination core.  Equation~(\ref{Qoutsidepreliminary}) then reduces to
\begin{align}
Q_{ij}=s_\mathrm{bulk}\biggl[&\frac{1}{4}\delta_{ij}-\frac{3}{4}\Omega_i \Omega_j
+\frac{3\hat{\mathbf{v}}\cdot\mathbf{r}}{4\rho}(m_i m_j -m'_i m'_j)\nonumber\\
&+\frac{3\hat{\mathbf{v}}'\cdot\mathbf{r}}{4\rho}(m_i m'_j +m'_i m_j)\biggr], \textrm{ for } \rho>\xi.
\label{Qoutside}
\end{align}

For the simplest model of the $Q_{ij}$ tensor inside the disclination core, we just make a linear interpolation of Eq.~(\ref{Qoutside}), which gives
\begin{align}
Q_{ij}=s_\mathrm{bulk}\biggl[&\frac{1}{4}\delta_{ij}-\frac{3}{4}\Omega_i \Omega_j
+\frac{3\hat{\mathbf{v}}\cdot\mathbf{r}}{4\xi}(m_i m_j -m'_i m'_j)\nonumber\\
&+\frac{3\hat{\mathbf{v}}'\cdot\mathbf{r}}{4\xi}(m_i m'_j +m'_i m_j)\biggr], \textrm{ for } \rho<\xi.
\label{Qinside}
\end{align}
This interpolated tensor exactly matches Eq.~(\ref{Qoutside}) at $\rho=\xi$.  The eigenvalues of this interpolated tensor are
\begin{equation}
s_1 =s_\mathrm{bulk}\left[\frac{1}{4}+\frac{3\rho}{4\xi}\right], \quad
s_2 =s_\mathrm{bulk}\left[\frac{1}{4}-\frac{3\rho}{4\xi}\right], \quad
s_3 =-\frac{s_\mathrm{bulk}}{2}.
\end{equation}
Those eigenvalues are shown by the solid lines in Fig.~3.  The tensor is biaxial (with three distinct eigenvalues) in most of the defect core, and it is uniaxial with a negative order parameter at $\rho=0$.  

\begin{figure}
\includegraphics[width=\columnwidth]{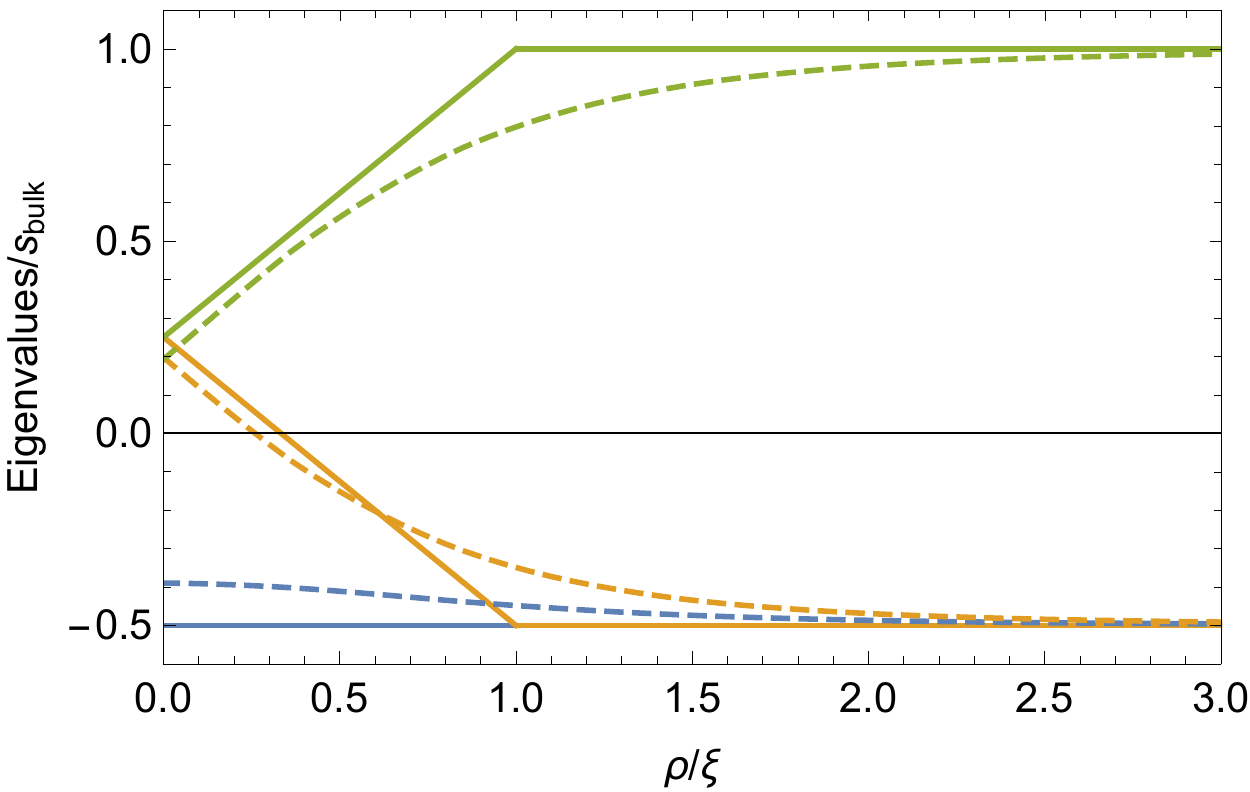}
\caption{Plots of the three eigenvalues of the nematic order tensor $Q_{ij}$ (normalized by the bulk order parameter $s_\mathrm{bulk}$) as functions of the radial coordinate $\rho$ outward from the disclination core (normalized by the core radius $\xi$).  Solid lines are the linear interpolation of Eqs.~(\ref{Qoutside}--\ref{Qinside}).  Dashed lines are a numerical solution using openQmin software.} 
\end{figure}

For a more precise model of $Q_{ij}$ inside the disclination core, one must minimize the free energy by solving the Euler-Lagrange equations, as was done by Schopohl and Sluckin.\cite{Schopohl1987}  One example of a numerical solution is shown by the dashed lines in Fig.~3.  We can see that it is generally similar to the simple linear interpolation, except that it smooths out the discontinuous derivative at $\rho=\xi$.  Hence, we will just use the linear interpolation of Eqs.~(\ref{Qoutside}--\ref{Qinside}) for the rest of this analysis.

Now we can use the $Q_{ij}$ tensor field to characterize the orientational features of the disclination.  The most fundamental feature of the disclination is the tensor at the center of the disclination core.  Evaluating Eq.~(\ref{Qinside}) at $\rho=0$ gives
\begin{equation}
Q_{ij}=s_\mathrm{bulk}\left[\frac{1}{4}\delta_{ij}-\frac{3}{4}\Omega_i \Omega_j\right].
\label{pancake}
\end{equation}
This tensor is uniaxial, with a principal eigenvalue of $-s_\mathrm{bulk}/2$ corresponding to the $\hat{\pmb{\Omega}}$ direction, and two eigenvalues of $+s_\mathrm{bulk}/4$ corresponding to the two orthogonal directions.  It can be visualized as an oblate ellipsoid (or pancake) with its short axis along $\hat{\pmb{\Omega}}$.  Its long axes are in the $(\hat{\mathbf{m}},\hat{\mathbf{m}}')$ plane, i.e.\ the plane of director rotation around the disclination.  Hence, this central $Q_{ij}$ tensor identifies the $\hat{\pmb{\Omega}}$ axis and the director plane.

For further orientational features, we must consider derivatives of $Q_{ij}$, evaluated inside the disclination core.  We can construct a vector orientation from the divergence of the tensor field,
\begin{align}
\partial_i Q_{ij} &= s_\mathrm{bulk}\left[\frac{3v_i}{4\xi}(m_i m_j -m'_i m'_j)+\frac{3v'_i}{4\xi}(m_i m'_j +m'_i m_j)\right]\nonumber\\
&= \frac{3s_\mathrm{bulk}}{2\xi}p_j.
\label{divQ}
\end{align}
Hence, this divergence of $Q_{ij}$ \emph{inside} the core provides a way to calculate the $\mathbf{p}$ vector associated with the disclination, defined in Eq.~(\ref{pdefinition}).  It carries the same orientational information as the average splay and bend vectors \emph{outside} the core, which were discussed in the previous section.

Beyond the divergence, we can also calculate the full third-rank gradient tensor $\partial_k Q_{ij}$ consisting of all the first derivatives, evaluated inside the disclination core,
\begin{equation}
\partial_k Q_{ij} = s_\mathrm{bulk}\left[\frac{3v_k}{4\xi}(m_i m_j -m'_i m'_j)+\frac{3v'_k}{4\xi}(m_i m'_j +m'_i m_j)\right].
\label{gradQ}
\end{equation}
Like the third-rank tensor $\overline{\Delta_{ij}n_k}$ in the previous section, this tensor breaks up into three components based on how they transform under rotations about $\hat{\pmb{\Omega}}$, with rotation matrix $R_{ij}(\phi)$.  Specifically, we have
\begin{equation}
\partial_k Q_{ij}=\partial_k Q_{ij}^{(1)}+\partial_k Q_{ij}^{(2)}+\partial_k Q_{ij}^{(3)},
\end{equation}
where
\begin{equation}
\partial_k Q_{ij}^{(l)}=\frac{1}{\pi}\int_{-\pi}^{\pi}d\phi\cos(l\phi)R_{ii'}(\phi)R_{jj'}(\phi)R_{kk'}(\phi)\partial_{k'} Q_{i'j'},
\label{integralconstruction}
\end{equation}
for $l=1$, $2$, and $3$.  These three components can be written as
\begin{align}
&\partial_k Q_{ij}^{(1)}=\frac{3s_\mathrm{bulk}}{4\xi}\left[p_i (\delta_{jk}-\Omega_j \Omega_k)
+p_j (\delta_{ik}-\Omega_i \Omega_k)-p_k (\delta_{ij}-\Omega_i \Omega_j)\right],\nonumber\\
&\partial_k Q_{ij}^{(2)}=\frac{3s_\mathrm{bulk}}{4\xi}\Omega_k D_{ij},\qquad
\partial_k Q_{ij}^{(3)}=\frac{3s_\mathrm{bulk}}{4\xi}T_{ijk},
\label{Qcomponents}
\end{align}
where $D_{ij}$ and $T_{ijk}$ are defined in Eqs.~(\ref{Ddefinition}) and~(\ref{Tdefinition}), respectively.  Hence, this gradient tensor carries the same information as $\overline{\Delta_{ij}n_k}$.  In particular, $\partial_k Q_{ij}^{(1)}$ provides information about the one-fold symmetric orientation of a wedge $+1/2$ disclination, $\partial_k Q_{ij}^{(2)}$ provides information about the two-fold symmetric orientation of a twist disclination, and $\partial_k Q_{ij}^{(3)}$ provides information about the three-fold symmetric orientation of a wedge $-1/2$ disclination.

In this formalism, the sign of the tangent vector $\hat{\mathbf{t}}$ is ambiguous, because one can move in either direction along the disclination line.  The sign of the rotation vector $\hat{\pmb{\Omega}}$ is also ambiguous, because Eq.~(\ref{pancake}) only defines the tensor $\Omega_i \Omega_j$.  However, these two choices of sign are related; the sign of $\hat{\mathbf{t}}$ determines the sign of $\hat{\pmb{\Omega}}$, or vice versa.  To see that relation, we can use Eqs.~(\ref{divQ}) and~(\ref{gradQ}) to derive
\begin{align}
\label{gradQmagnitude}
&(\partial_k Q_{ij})(\partial_k Q_{ij})=\left[\frac{3s_\mathrm{bulk}}{2\xi}\right]^2,\\
&(\partial_i Q_{ij})(\partial_k Q_{kj})=\left[\frac{3s_\mathrm{bulk}}{2\xi}\right]^2 \left[\frac{1+\cos\beta}{2}\right]^2
.
\end{align}
By taking the ratio of those expressions, we obtain the dot product
\begin{equation}
\hat{\pmb{\Omega}}\cdot\hat{\mathbf{t}}=\cos\beta=-1+2|\mathbf{p}|=
-1+2\left[\frac{(\partial_i Q_{ij})(\partial_k Q_{kj})}{(\partial_k Q_{ij})(\partial_k Q_{ij})}\right]^{1/2}.
\label{Omegadott}
\end{equation}
Hence, once the sign of $\hat{\mathbf{t}}$ is chosen, there is no further ambiguity in $\hat{\pmb{\Omega}}$.

\section{Geometric analysis of simulations}

\begin{figure*}
(a)\includegraphics[width=\columnwidth]{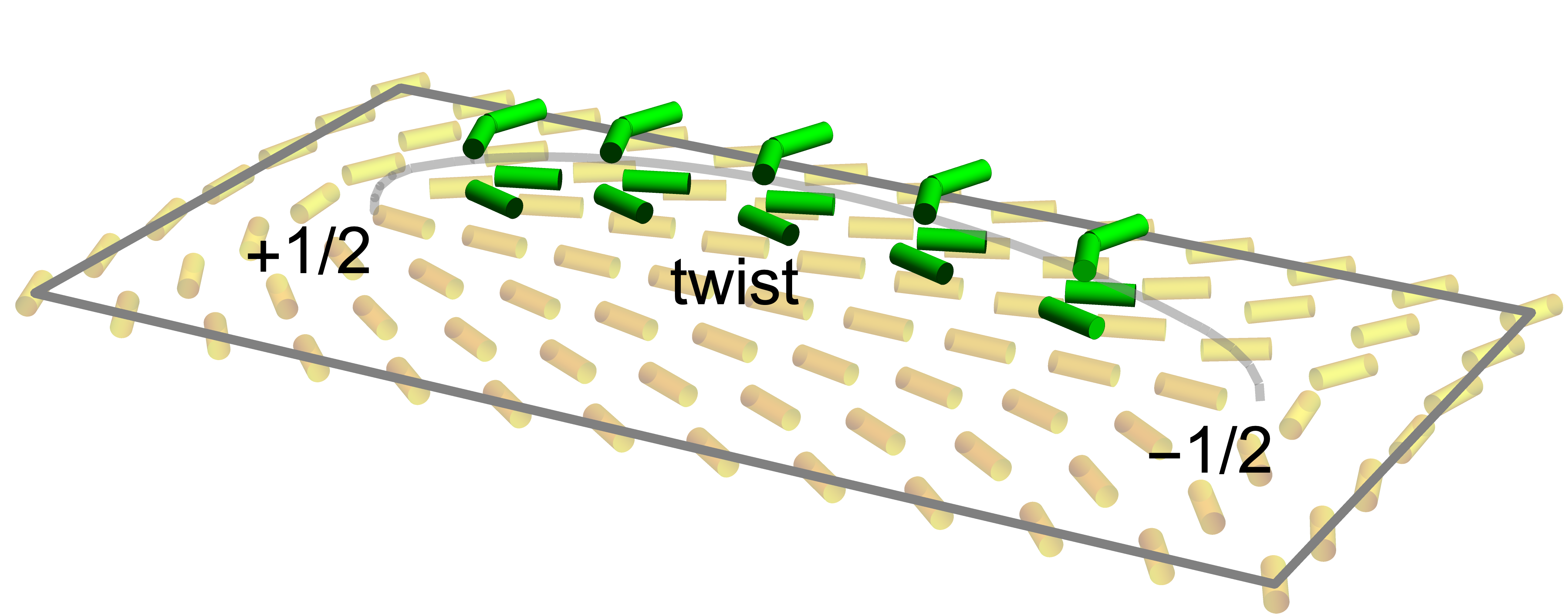}\includegraphics[width=\columnwidth]{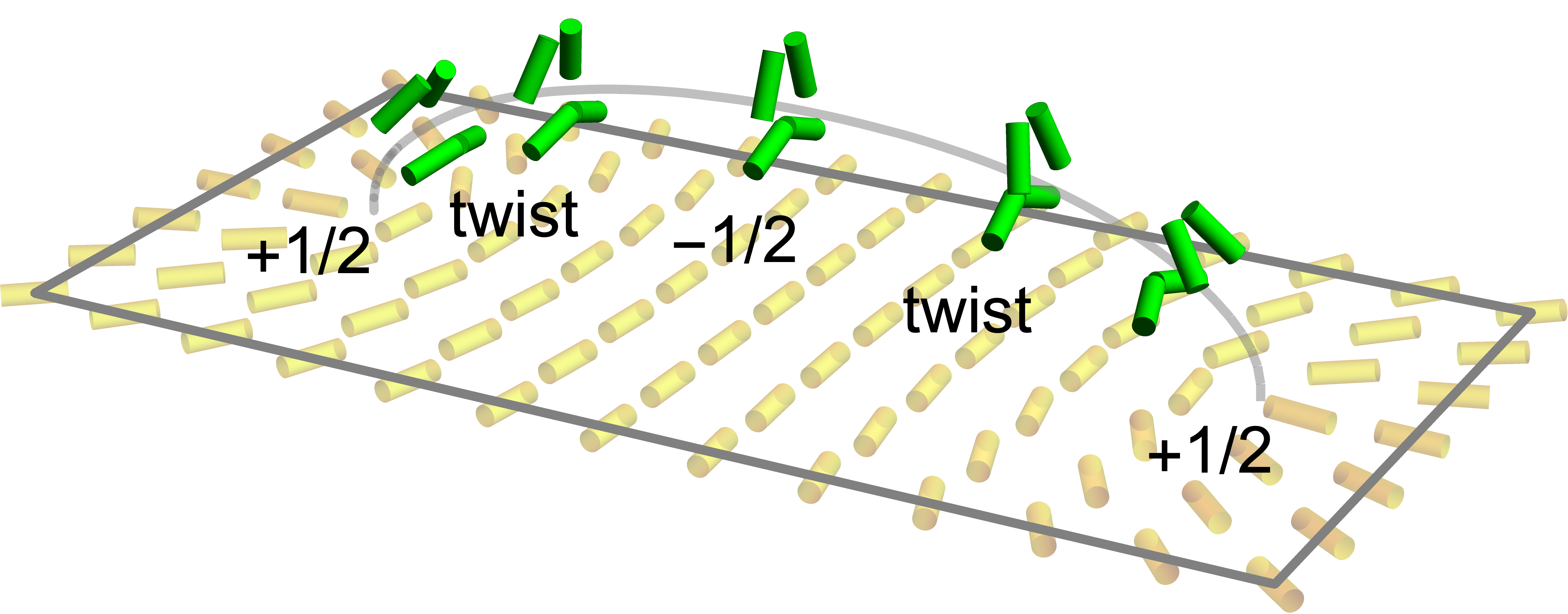}
(b)\includegraphics[width=\columnwidth]{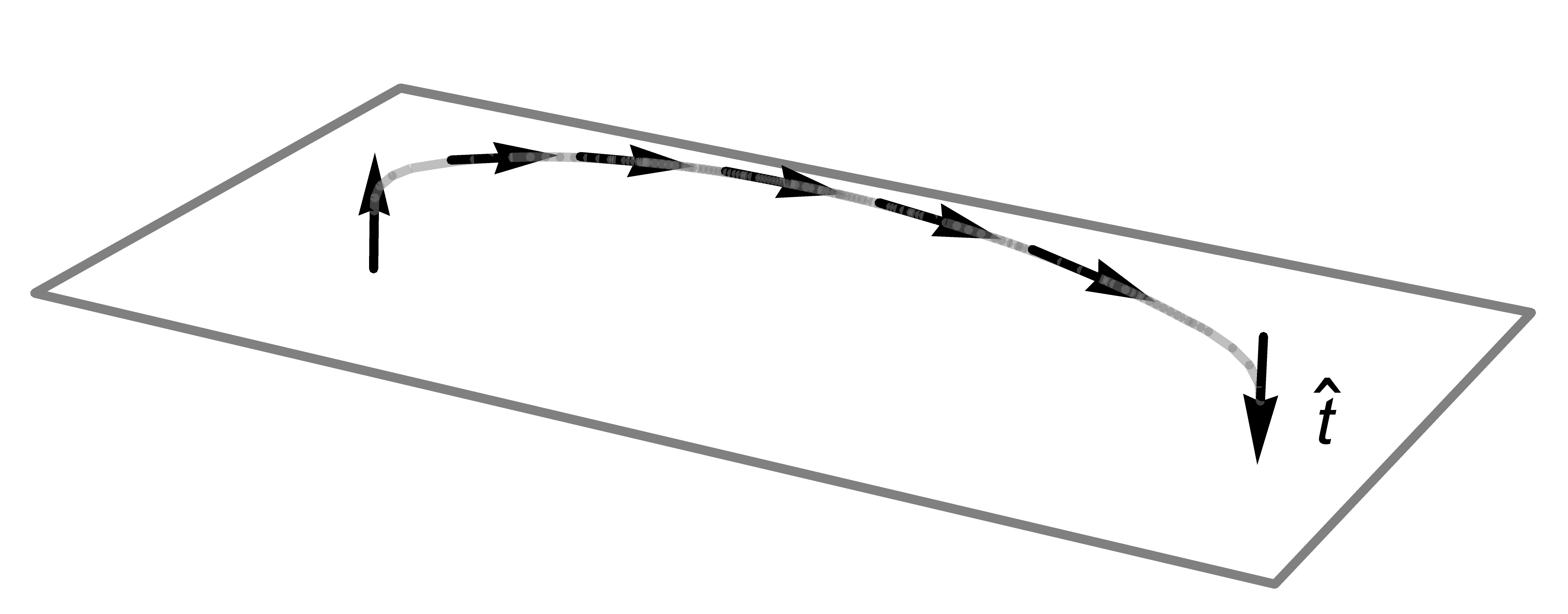}\includegraphics[width=\columnwidth]{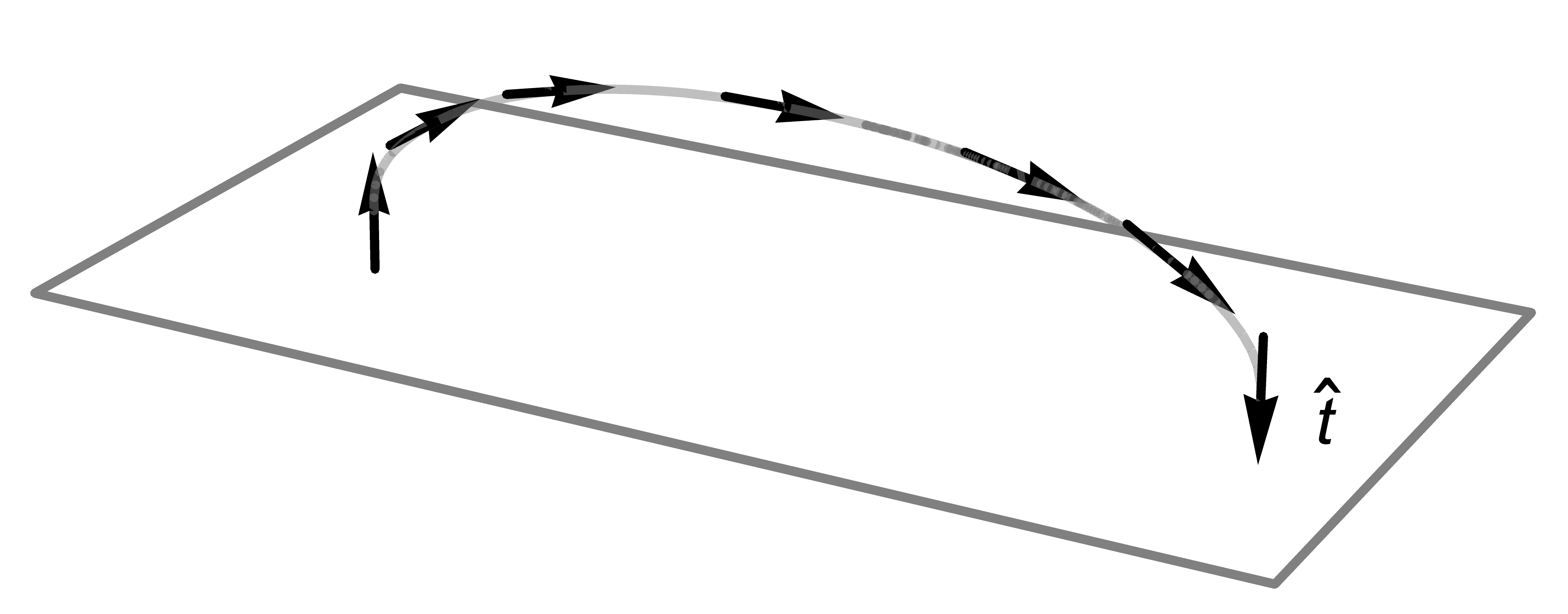}
(c)\includegraphics[width=\columnwidth]{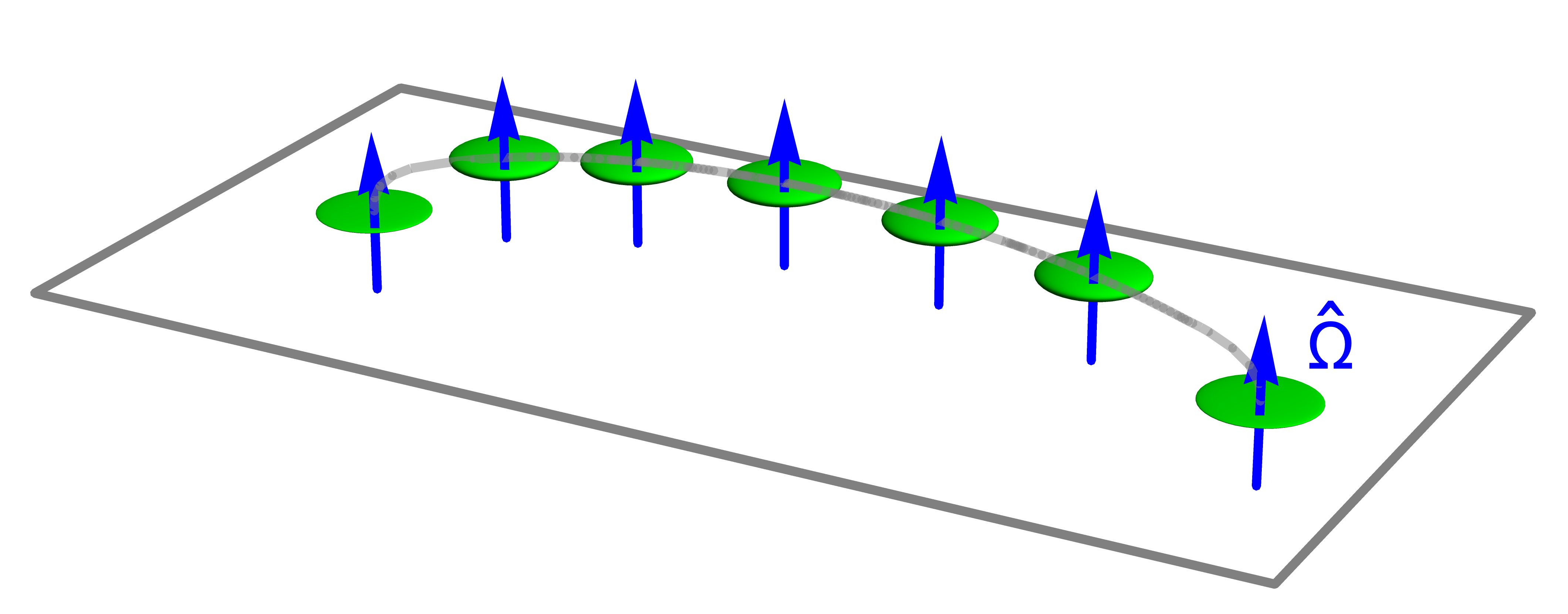}\includegraphics[width=\columnwidth]{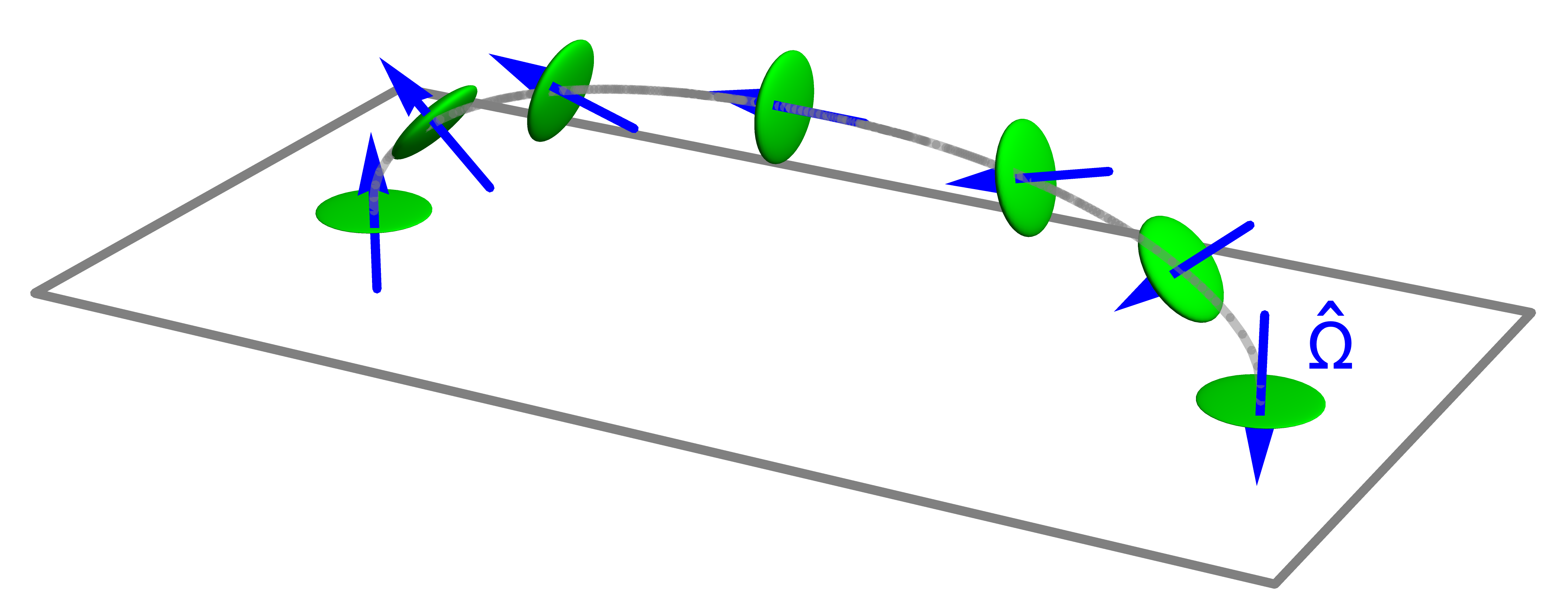}
(d)\includegraphics[width=\columnwidth]{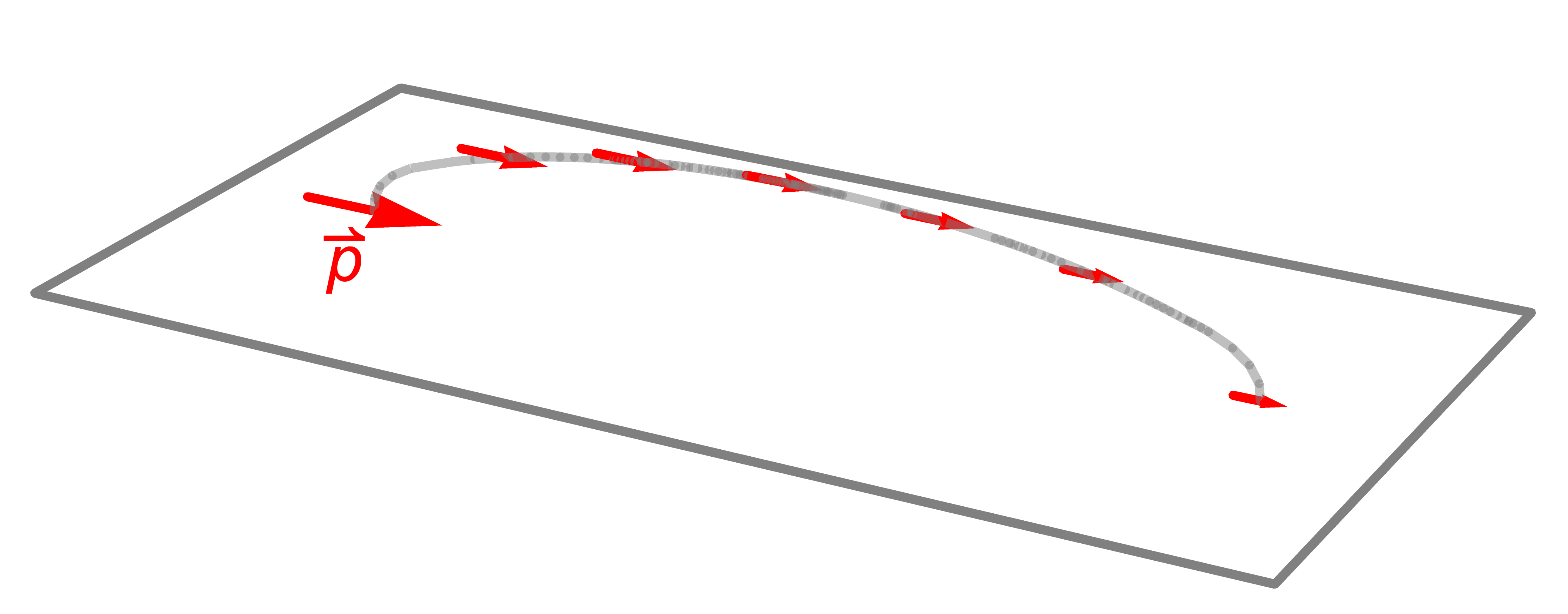}\includegraphics[width=\columnwidth]{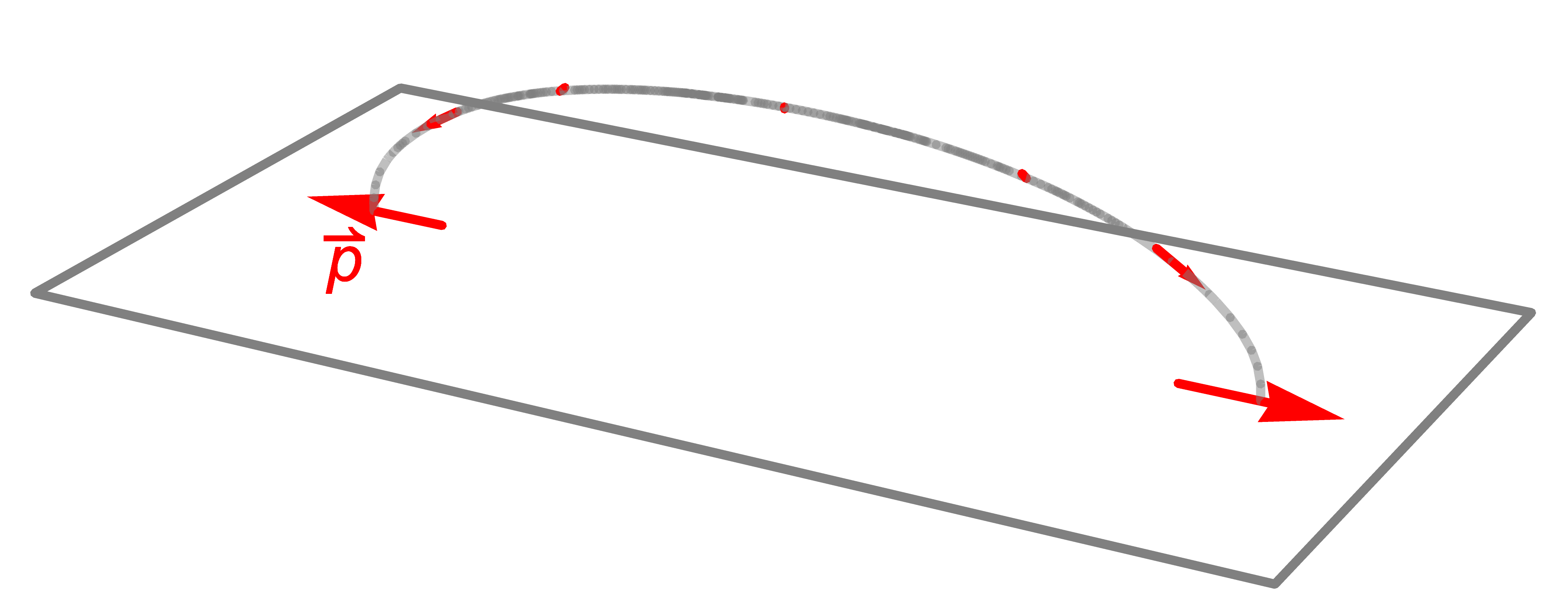}
(e)\includegraphics[width=\columnwidth]{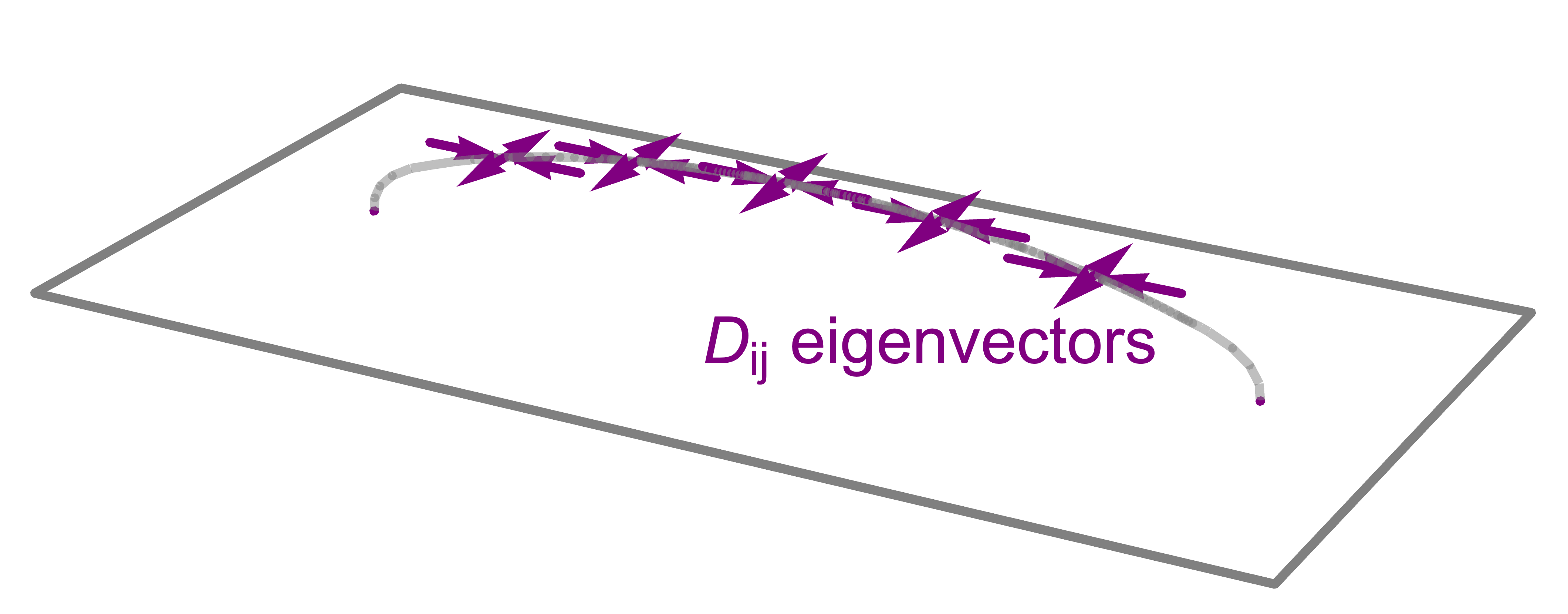}\includegraphics[width=\columnwidth]{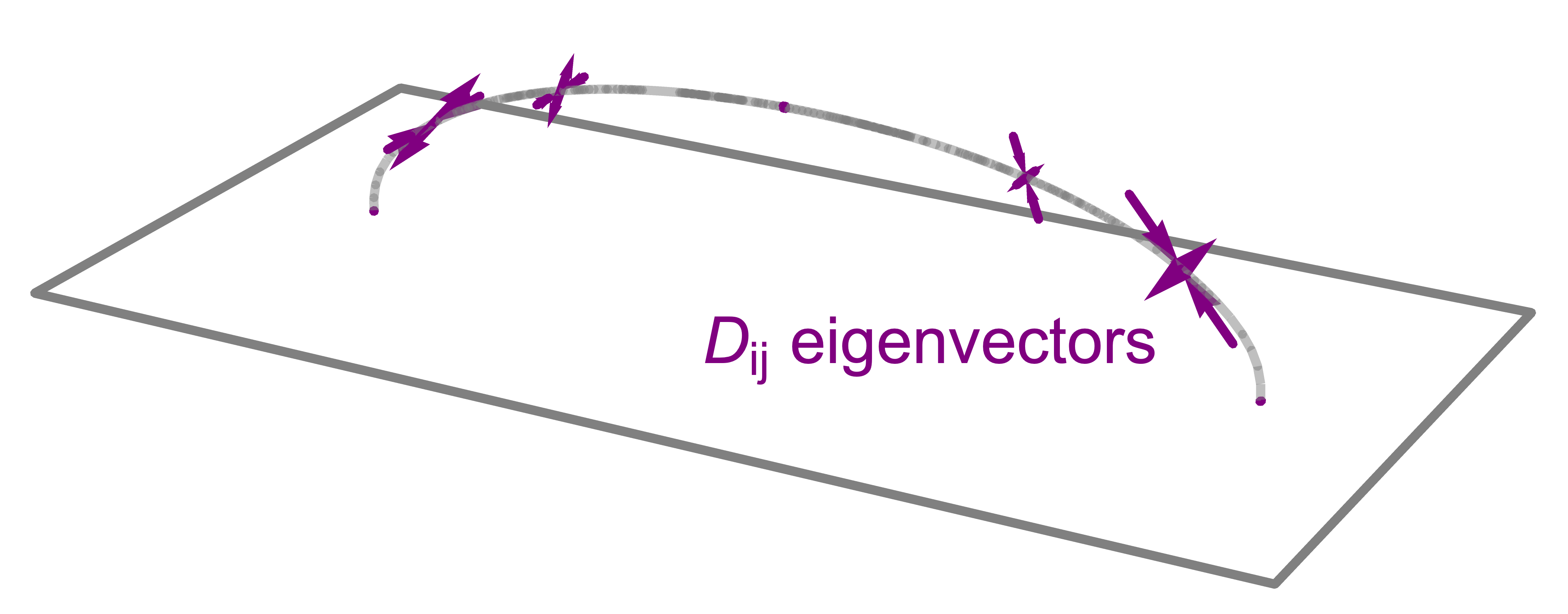}
(f)\includegraphics[width=\columnwidth]{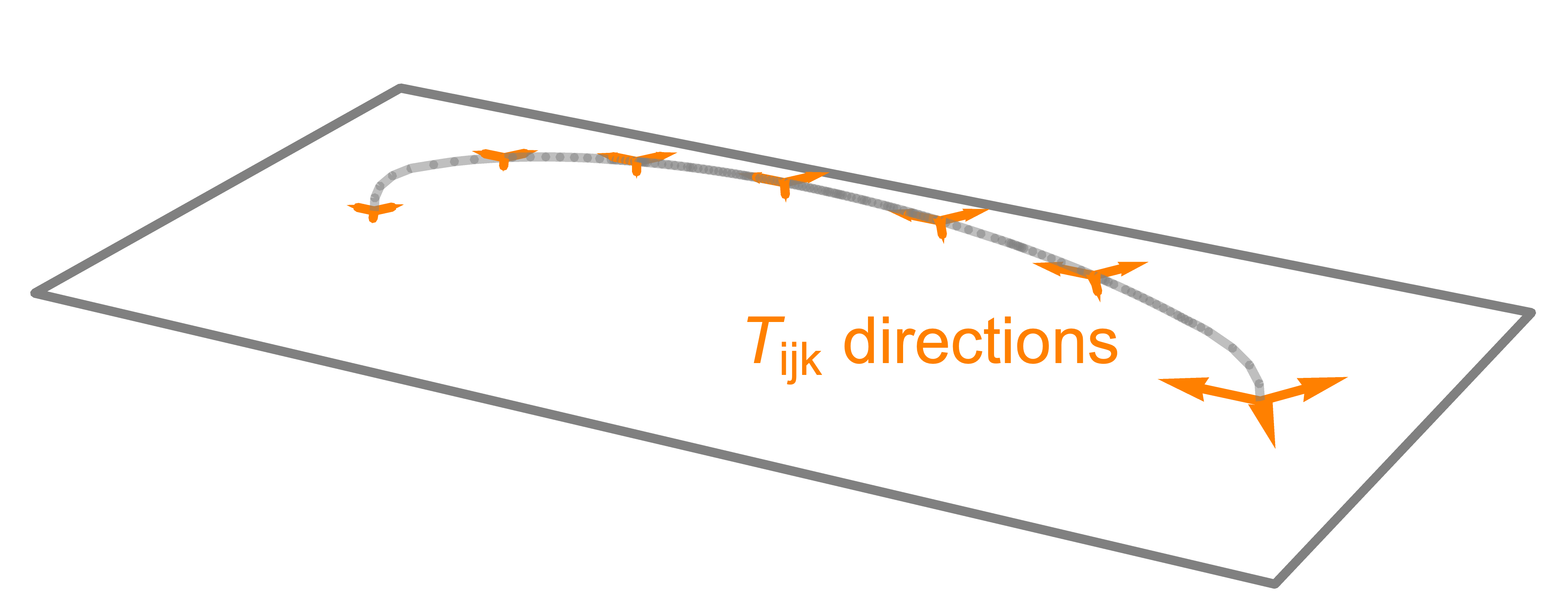}\includegraphics[width=\columnwidth]{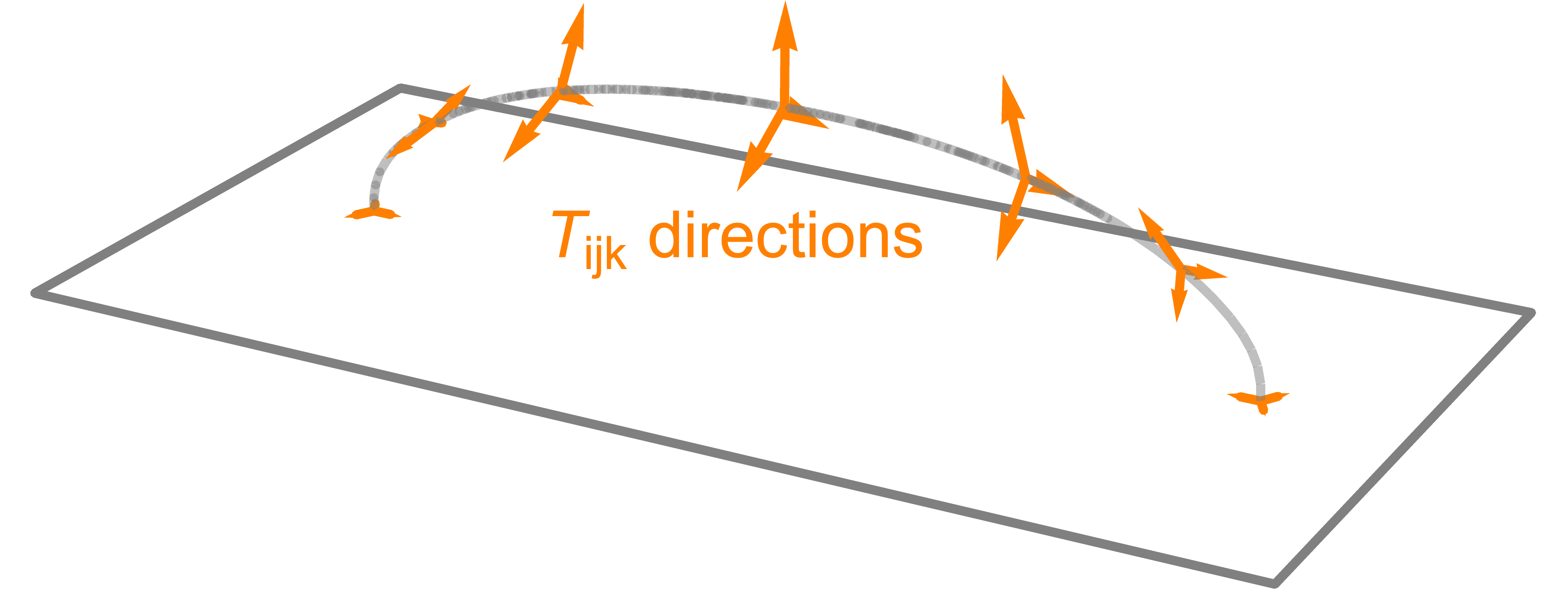}
\caption{Two simulations of disclination lines in a 3D nematic liquid crystal, which connect surface disclinations on the bottom substrate.  Each column shows six alternative visualizations of the same simulation.  (a)~Cylinders representing the director field (green around the disclination line in the bulk, yellow on the bottom substrate).  (b)~Tangent vector $\hat{\mathbf{t}}$ along the disclination line.  (c)~Oblate ellipsoids representing the $Q_{ij}$ tensor at the center of the disclination core, and arrows representing the rotation vector $\hat{\pmb{\Omega}}$.  (d)~Vector $\mathbf{p}$ showing the one-fold symmetric component of disclination orientation, which is greatest wherever a disclination has $+1/2$ wedge structure.  (e)~Eigenvectors of the $D_{ij}$ tensor showing the two-fold symmetric component of disclination orientation, which is greatest wherever a disclination has twist structure.  (f)~Triad associated with $T_{ijk}$, showing the three-fold symmetric component of disclination orientation, which is greatest wherever a disclination has $-1/2$ wedge structure.}
\end{figure*}

For specific examples of these geometric constructions, we perform two simulations of disclination lines in 3D nematic liquid crystals.  In the first simulation, shown in the left column of Fig.~4, the bottom substrate is patterned with a $+1/2$ and a $-1/2$ surface disclination.  The top and side surfaces have free boundary conditions.  Responding to these surfaces, the bulk liquid crystal forms a disclination line that connects the two surface disclinations.  In the second simulation, shown in the right column of Fig.~4, the bottom substrate is patterned with two $+1/2$ surface disclinations.  The top surface has homeotropic anchoring, and the side surfaces are free.  Again, in response to these boundary conditions, the bulk liquid crystal forms a disclination line connecting the two surface disclinations.

These simulations are similar to other simulations that have been published by our group,\cite{Afghah2018,Ferris2020} but they are new simulations.  They are done using the software package openQmin, created by Sussman and Beller.\cite{Sussman2019}.  This software minimizes the Landau-de Gennes free energy, expressed in terms of the nematic order tensor $Q_{ij}$, discretized on a cubic lattice.  We use the default scaled Landau-de Gennes parameters $A=-1$, $B=-12.3$, $C=10.1$, and $L=4.64$, which give a disclination line core radius of $\xi=(L/A)^{1/2}=2.2$ scaled units.  In both simulations, the distance between the surface disclinations is 45 scaled units, and the system size is $102\times102\times22$ scaled units.  To analyze the simulation results, we use the following procedure:

First, we interpolate a smooth tensor field $Q_{ij}(\mathbf{r})$ between the sites of the simulation lattice, using the interpolation function of Mathematica.  We find the disclination line by searching for local minima of $Q_{ij}Q_{ij}$, and connect minima by a smooth curve.  Outside the disclination core, the nematic order tensor is uniaxial with a positive order parameter $s_\mathrm{bulk}=0.53$.  Figure~4(a) shows visualizations of the director field, i.~e.\ the eigenvector of $Q_{ij}$ corresponding to the positive eigenvalue $s_\mathrm{bulk}$.  The yellow cylinders represent the fixed director field on the bottom substrate, while the green cylinders represent the simulated director field in the interior.  Green cylinders are only shown for selected positions around the disclination line; they are omitted everywhere else for simplicity.

From these visualizations, we can see that both disclination lines change their structure along the length of the line.  In the first simulation, the disclination begins as a $+1/2$ wedge disclination at one surface anchoring point, then changes to a twist disclination at the top of the arch, then becomes a $-1/2$ wedge disclination at the other surface anchoring point.  In the second simulation, the disclination begins as a $+1/2$ wedge, then changes to a twist, a $-1/2$ wedge at the top of the arch, again a twist, and ends as a $+1/2$ wedge once again.

Second, we construct the unit tangent vector $\hat{\mathbf{t}}$ along each disclination line.  These unit vectors are shown by the black arrows in Fig.~4(b).  As discussed previously, the overall sign of $\hat{\mathbf{t}}$ is ambiguous.  For each simulation, we make an arbitrary choice of which end of the disclination is the ``beginning'' or the ``end,'' and this choice determines the sign of $\hat{\mathbf{t}}$.

Third, we calculate the tensor $Q_{ij}$ at the center of each disclination core, as a function of position along the line.  As shown in Fig.~3, at the center of the disclination core, the tensor $Q_{ij}$ has one negative eigenvalue of approximately $-\frac{1}{2}s_\mathrm{bulk}$ and two degenerate positive eigenvalues of approximately $+\frac{1}{4}s_\mathrm{bulk}$.  Hence, it can be represented by an oblate ellipsoid (pancake), as shown by the green ellipsoids in Fig.~4(c).  This ellipsoid identifies the local director plane at that point along the disclination.  In the first simulation, this director plane is always horizontal.  In the second simulation, this director plane rotates through $180^\circ$ from the beginning to the end of the disclination.

At the center of the disclination core, the eigenvector corresponding to the negative eigenvalue is $\pm\hat{\pmb{\Omega}}$.  It is shown by the blue arrows in Fig.~4(c).  We can see that this vector is normal to the director plane given by the ellipsoid.  To resolve the sign of $\pm\hat{\pmb{\Omega}}$, we calculate the dot product with the $\hat{\mathbf{t}}$, and choose the sign that gives the best agreement with Eq.~(\ref{Omegadott}).  The agreement is not exact because the equation is derived with the assumption of a straight disclination line, while the simulated disclinations are curved, but it is generally close enough to determine the sign.  In the regions where this criterion fails because $\hat{\pmb{\Omega}}\cdot\hat{\mathbf{t}}\approx0$, we determine the sign by requiring that $\hat{\pmb{\Omega}}$ vary smoothly along the disclination line.

Fourth, we calculate the gradient $\partial_k Q_{ij}$ and the divergence $\partial_i Q_{ij}$ by taking derivatives of the interpolated tensor field.  We then determine the vector $\mathbf{p}$ from the divergence using Eq.~(\ref{divQ}), normalized by $(3s_\mathrm{bulk})/(2\xi)$ calculated from Eq.~(\ref{gradQmagnitude}).  The results for $\mathbf{p}$ are given by the red arrows in Fig.~4(d).  The magnitude $|\mathbf{p}|\approx(1+\cos\beta)/2$ is largest wherever a disclination is a $+1/2$ wedge, and it is smaller wherever a disclination has a twist or $-1/2$ wedge structure.  (It does not exactly vanish at the $-1/2$ wedge in the first simulation, again because the theory is derived with the assumption of a straight disclination, while the simulated disclinations are curved.)  Wherever a disclination is a $+1/2$ wedge, the orientation of $\mathbf{p}$ matches the 2D orientation on the bottom substrate, as defined in Fig.~1.  Hence, it provides a consistent description of the one-fold symmetric aspect of disclination orientation.

Fifth, we construct the tensor $D_{ij}$, as given in Eq.~(\ref{Qcomponents}).  This construction is most conveniently done by contracting with the $\hat{\pmb{\Omega}}$ vector to obtain $\Omega_k \partial_k Q_{ij} = [(3s_\mathrm{bulk})/(4\xi)] D_{ij}$, normalized by $(3s_\mathrm{bulk})/(4\xi)$ from Eq.~(\ref{gradQmagnitude}).  The tensor $D_{ij}$ has eigenvalues of approximately $\pm\sin\beta$ and $0$.  In Fig.~4(e), we represent the eigenvector corresponding to $+\sin\beta$ as a double purple arrow going outward from the disclination line, and the eigenvector corresponding to $-\sin\beta$ as a double purple arrow going inward.  (The eigenvector corresponding to $0$ is just $\hat{\pmb{\Omega}}$, which was already shown in Fig.~4(c).)  We can see that the eigenvectors represent the two-fold symmetric orientation of a twist disclination, which has no analogue in 2D.  In the figure, the length of the eigenvectors is drawn proportional to the eigenvalue $\sin\beta$.  It is longest wherever a disclination has a twist structure, and vanishes wherever the disclination is a $\pm1/2$ wedge.  In the second simulation, there is a reversal between inward and outward arrows on the two sides of the arch.  This reversal occurs because the sign of $D_{ij}$ depends on the sign of $\hat{\pmb{\Omega}}$, which depends on the sign of $\hat{\mathbf{t}}$, and that sign is necessarily defined upward on one side of the arch and downward on the other.

Sixth, we determine the third-rank tensor $T_{ijk}$ by using the integral construction of Eq.~(\ref{integralconstruction}) with $l=3$.  This quantity is a completely symmetric tensor, which represents three-fold orientational order in the plane perpendicular to $\hat{\pmb{\Omega}}$.  To extract the orientation, we use the same procedure developed to analyze $-1/2$ disclinations in 2D.\cite{Tang2017}  We construct a test vector $\hat{\mathbf{g}}=\hat{\mathbf{m}}\cos\phi+\hat{\mathbf{m}}'\sin\phi$, and use it to define the scalar $f(\phi)=T_{ijk}g_i g_j g_k$.  We then search for the maxima of $f(\phi)$, which define three equivalent values separated by $120^\circ$ from each other, and hence a triad of vectors $\hat{\mathbf{g}}$.  This triad is represented by the orange arrows in Fig.~4(f).  The length of the arrows is drawn proportional to the magnitude $\frac{1}{2}[T_{ijk}T_{ijk}]^{1/2}\approx(1-\cos\beta)/2$.  It is longest wherever a disclination has a $-1/2$ wedge structure, and is reduced wherever a disclination has a twist or $+1/2$ wedge structure.  Wherever a disclination is a $-1/2$ wedge, the orientation of the triad matches the 2D concept of $-1/2$ disclination orientation, as defined in Fig.~1.  Hence, it provides a consistent description of the three-fold symmetric aspect of disclination orientation.

This computational work shows that the disclination in the first simulation has a much lower free energy than the disclination in the second simulation, because the director plane and the $\hat{\pmb{\Omega}}$ vector are constant in the first simulation, but they rotate through $180^\circ$ in the second simulation.  The second disclination can only form because of the conflict between planar anchoring on the bottom and homeotropic anchoring on the top, which forces the system into a high energy state.  Hence, one general lesson is that topology does not require $\hat{\pmb{\Omega}}$ to be constant along the length of a disclination, but energy strongly favors a constant $\hat{\pmb{\Omega}}$ along a disclination line.

\section{Peach-Koehler force}

For the rest of this paper, we will use the geometric formalism established above to characterize forces acting on disclination lines in 3D nematic liquid crystals.  In this section, we consider the liquid-crystal version of the Peach-Koehler force.

The Peach-Koehler force is a well-known force acting on dislocation lines in crystalline solids.\cite{Peach1950,Lubarda2019}  When a crystalline solid is subjected to an external stress $\pmb{\sigma}$, any dislocation line in that solid experiences a force per unit length of $\mathbf{f}_\mathrm{PK}=(\pmb{\sigma}\cdot\mathbf{b})\times\hat{\mathbf{t}}$, where $\mathbf{b}$ is the Burgers vector and $\hat{\mathbf{t}}$ is the local unit tangent vector to the dislocation.  This force has important effects on the microstructural evolution and mechanical response of solids.

The concept of a Peach-Koehler force was applied to nematic liquid crystals many years ago by Kl\'{e}man.\cite{Kleman1983}  To our knowledge, it has only been mentioned in the nematic liquid crystal literature a few times since then.\cite{Eshelby1980,Kawasaki1985,Rey1990}  Here, we investigate some specific cases, in order to see how the Peach-Koehler force affects the behavior of disclination lines.

To begin, let us consider the simplified case of a liquid crystal in which the director field is always in the $(x,y)$ plane, and it only depends on $x$ and $y$, independent of $z$.  In that case, we can describe the director in terms of an angle field $\theta$, such that $\hat{\mathbf{n}}(x,y)=(\cos\theta(x,y),\sin\theta(x,y),0)$.  This effectively 2D liquid crystal can have disclinations with topological charges of $\pm1/2$.  Both types of disclinations extend vertically in the $z$-direction, and hence we can choose the tangent vector $\hat{\mathbf{t}}=\hat{\mathbf{z}}$.  For a $+1/2$ disclination, the rotation vector $\hat{\pmb{\Omega}}$ must be parallel to $\hat{\mathbf{t}}$, and hence $\hat{\pmb{\Omega}}=\hat{\mathbf{z}}$.  For a $-1/2$ disclination, the rotation vector $\hat{\pmb{\Omega}}$ must be antiparallel to $\hat{\mathbf{t}}$, and hence $\hat{\pmb{\Omega}}=-\hat{\mathbf{z}}$.

These two types of disclinations can be mapped onto \emph{screw dislocations} in a crystalline solid.\cite{Li1999,Selinger2000}  For the $+1/2$ disclination, the angle field $\theta$ increases by $\pi$ as one moves in a loop around the disclination.  This behavior maps onto a crystalline solid in which the displacement field $u_z$ increases by $b$ as one moves in a loop around a screw dislocation, and hence the Burgers vector is $\mathbf{b}=+b\hat{\mathbf{z}}$.  Likewise, for the $-1/2$ disclination, $\theta$ decreases by $\pi$ as one moves in a loop around the disclination.  This behavior maps onto a solid in which $u_z$ decreases by $b$ as one moves in a loop around a screw dislocation, and hence the Burgers vector is $\mathbf{b}=-b\hat{\mathbf{z}}$.  In both cases, the effective Burgers vector is
\begin{equation}
\mathbf{b}^\mathrm{eff}=\pi\hat{\pmb{\Omega}}.
\label{effectiveburgers}
\end{equation}
This effective Burgers vector is dimensionless, unlike the Burgers vector of a solid, which has dimensions of length.

It might seem surprising that a $\pm1/2$ wedge disclination in a nematic liquid crystal, which is not chiral, maps onto a screw dislocation in a crystalline solid, which is chiral.  However, this mapping is reasonable, because it depends on the angle $\theta$, which is defined with a particular handedness; it measures the angle of the director away from the $x$-axis in the counter-clockwise direction.

Now that we have a mapping of disclinations onto dislocations, we can express the mechanics of a nematic liquid crystal in a style analogous to the mechanics of a crystalline solid.  Because $\theta$ corresponds to the displacement $u_z$, derivatives of $\theta$ must correspond to the strain, which involves derivatives of the displacement.  Hence, we define the effective strain tensor as
\begin{equation}
\epsilon^\mathrm{eff}_{zi}=\partial_i \theta.
\label{effectivestrain}
\end{equation}
This effective strain tensor is not symmetric, unlike the conventional strain tensor, which is symmetrized between its indices.  Also, this effective strain tensor has dimensions of inverse length, unlike the conventional strain tensor, which is dimensionless.

In a solid, the elastic energy can be expressed in terms of the strain tensor, and the same is true for a nematic liquid crystal.  In the simplest model with equal Frank constants, the elastic free energy is
\begin{equation}
F=\frac{1}{2}K(\partial_i n_j)(\partial_i n_j)
=\frac{1}{2}K(\partial_i \theta)(\partial_i \theta)
=\frac{1}{2}K\epsilon^\mathrm{eff}_{zi}\epsilon^\mathrm{eff}_{zi}.
\end{equation}
Furthermore, in a solid, the stress can be defined as a derivative of the elastic energy with respect to strain.  Hence, for a liquid crystal, we define the effective stress tensor as
\begin{equation}
\sigma^\mathrm{eff}_{zi}=\frac{\partial F}{\partial\epsilon^\mathrm{eff}_{zi}}
=K\epsilon^\mathrm{eff}_{zi}
=K\partial_i \theta.
\label{effectivestress}
\end{equation}
This effective stress tensor is not symmetric and has dimensions of force/length or torque/area, unlike the conventional stress tensor, which has dimensions of force/area.

The physical interpretation of the effective stress tensor is the difference of torques across a cell, normalized by the area of the cell.  Suppose we apply a torque of $+\tau\hat{\mathbf{z}}$ on one surface of a cell at $y=+d/2$, and the opposite torque of $-\tau\hat{\mathbf{z}}$ on the opposite surface at $y=-d/2$.  This difference of torques can be expressed as an effective stress of $\sigma^\mathrm{eff}_{zy}=\tau/(L_x L_z)$, where $(L_x L_z)$ is the surface area of the cell in the $(x,z)$ plane.  For any small volume inside the liquid crystal, the total torque acting on $\theta$ is $\tau_z=(\partial_i \sigma^\mathrm{eff}_{zi})\Delta x\Delta y\Delta z = (K\nabla^2 \theta)\Delta x\Delta y\Delta z$.

This mapping leads to the nematic analogue of the Peach-Koehler force.  In solids, the Peach-Koehler force (per length) is normally written as $\mathbf{f}_\mathrm{PK}=(\pmb{\sigma}\cdot\mathbf{b})\times\hat{\mathbf{t}}$.  Of course, because the stress tensor is symmetric, it can equivalently be written as $\mathbf{f}_\mathrm{PK}=(\mathbf{b}\cdot\pmb{\sigma})\times\hat{\mathbf{t}}$.  In a nematic liquid crystal, the effective stress tensor defined above is not symmetric, and hence we must decide whether to contract the effective stress tensor with the effective Burgers vector as $\pmb{\sigma}^\mathrm{eff}\cdot\mathbf{b}^\mathrm{eff}$ or $\mathbf{b}^\mathrm{eff}\cdot\pmb{\sigma}^\mathrm{eff}$.  From Eqs.~(\ref{effectiveburgers}) and~(\ref{effectivestress}), we can see that $\pmb{\sigma}^\mathrm{eff}\cdot\mathbf{b}^\mathrm{eff}$ is trivially zero when $\theta$ depends only on $x$ and $y$, but $\mathbf{b}^\mathrm{eff}\cdot\pmb{\sigma}^\mathrm{eff}$ is nontrivial.  Hence, the nematic Peach-Koehler force (per length) can be constructed as
\begin{equation}
\mathbf{f}_\mathrm{PK}=(\pi\hat{\pmb{\Omega}}\cdot\pmb{\sigma}^\mathrm{eff})\times\hat{\mathbf{t}}.
\label{nematicPK}
\end{equation}
We should emphasize two features of this expression.  First, it has the correct dimensions of force/length, because the extra factor of length in $\pmb{\sigma}^\mathrm{eff}$ compensates for the missing factor of length in $\mathbf{b}^\mathrm{eff}=\pi\hat{\pmb{\Omega}}$.  Second, it is linear in both the rotation vector $\hat{\pmb{\Omega}}$ and the tangent vector $\hat{\mathbf{t}}$, and hence it keeps the same sign even if we change the sign of $\hat{\pmb{\Omega}}$ and $\hat{\mathbf{t}}$ simultaneously.  The sign of this force is not ambiguous.

\begin{figure}[t]
(a)\includegraphics[width=.9\columnwidth]{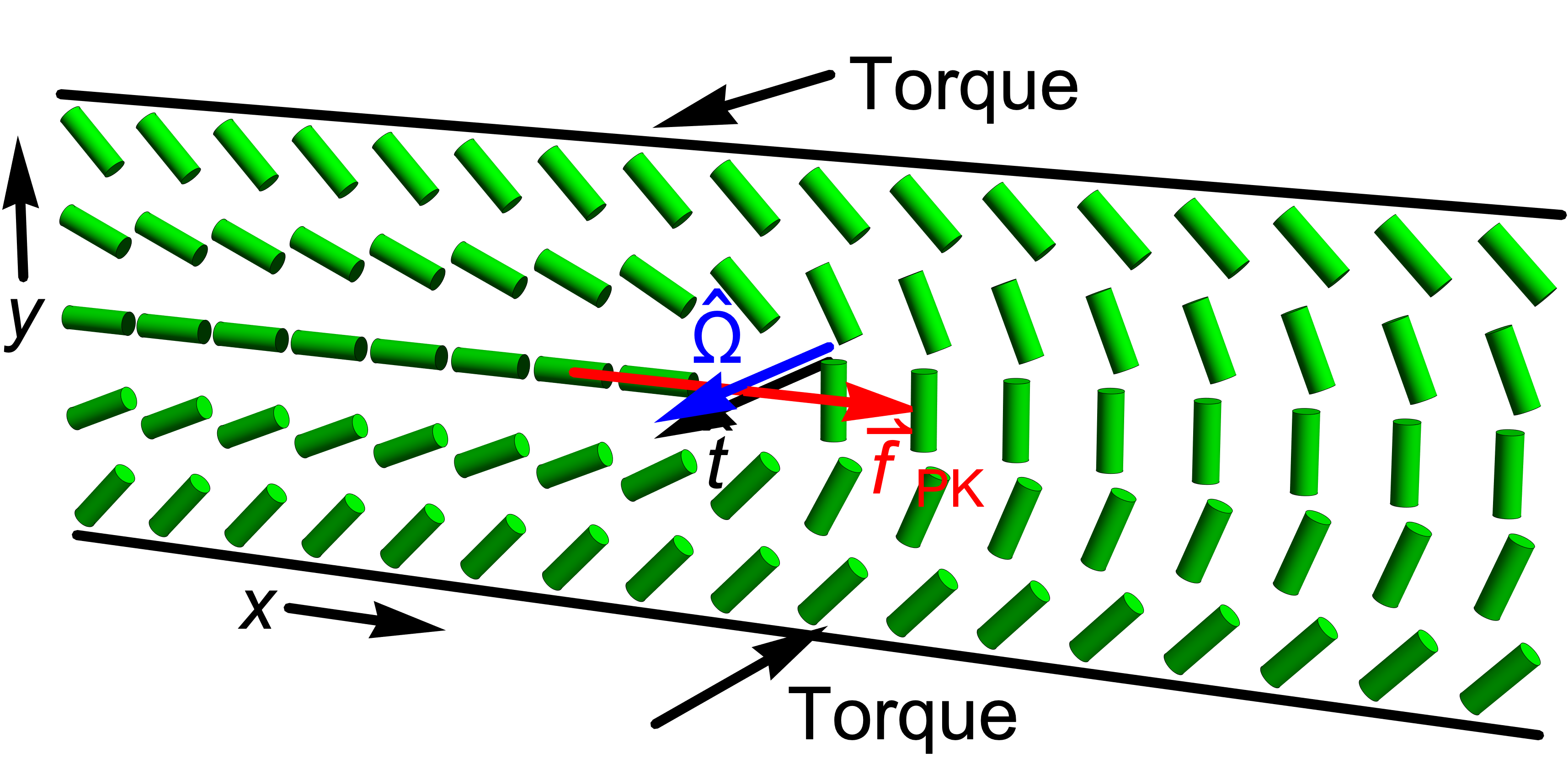}
(b)\includegraphics[width=.9\columnwidth]{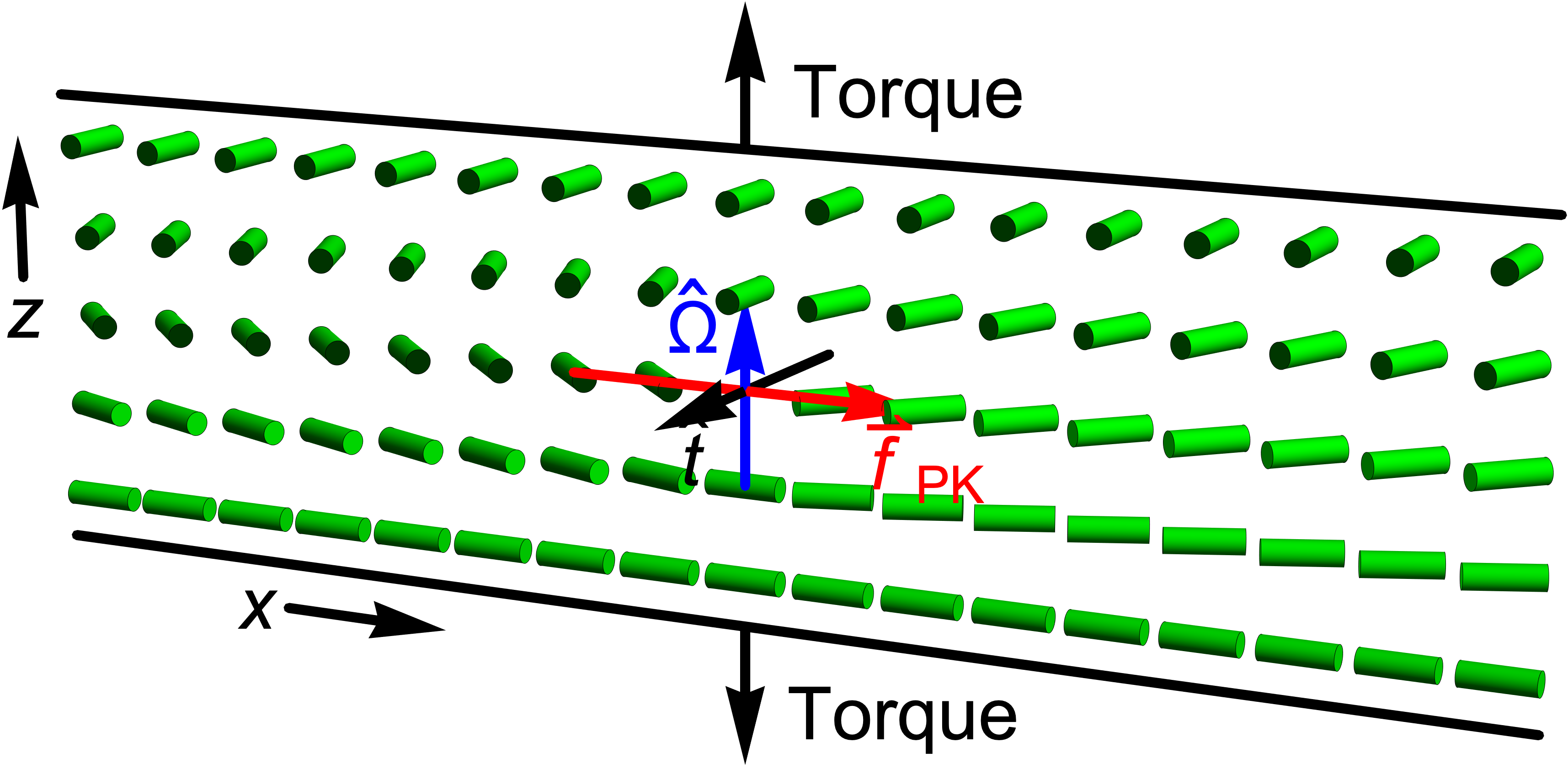}
(c)\includegraphics[width=.9\columnwidth]{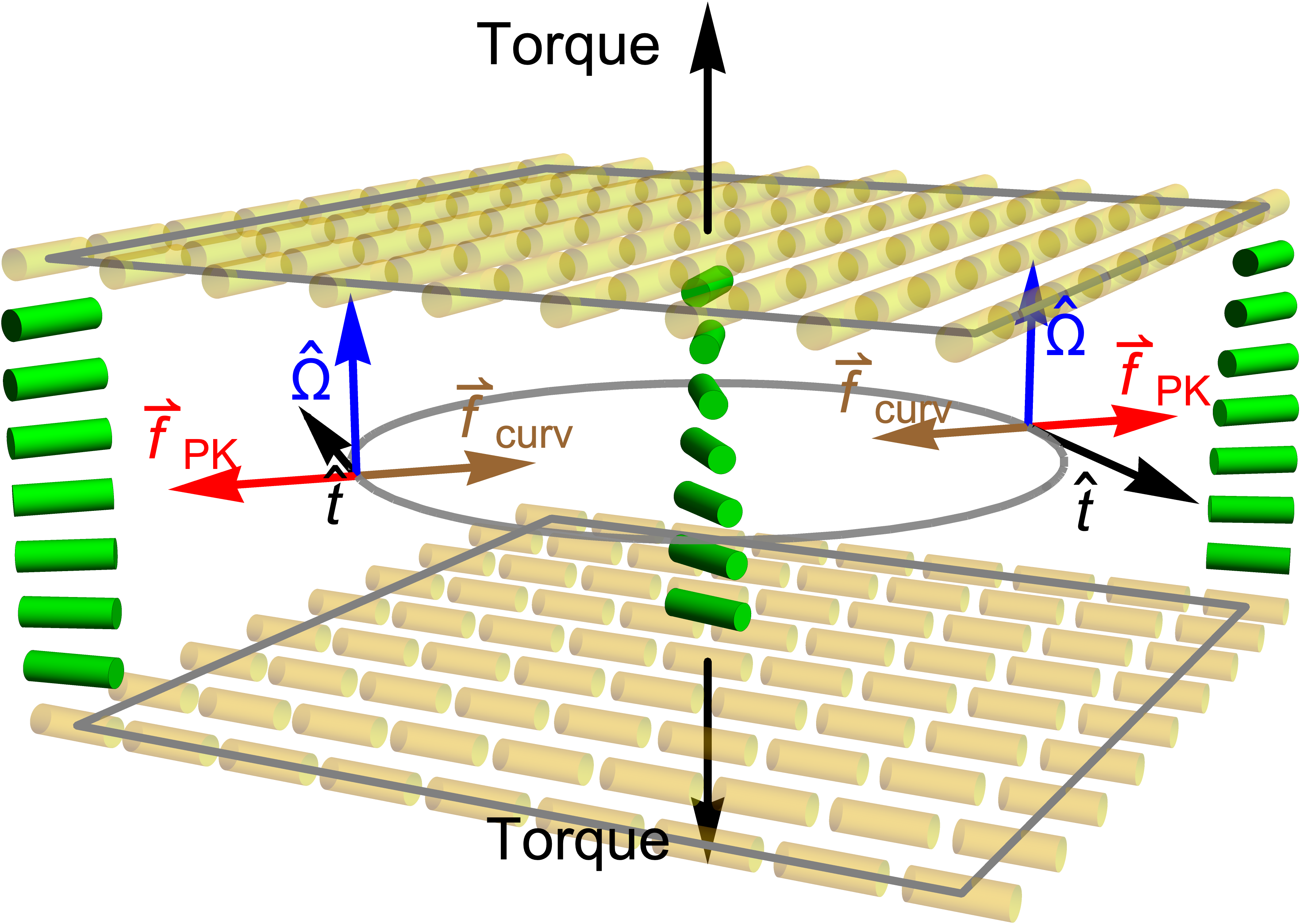}
\caption{Examples of the Peach-Koehler force.  (a)~Acting on a $+1/2$ wedge disclination line.  (b)~Acting on a twist disclination line.  (c)~Acting on a pure twist disclination loop.}
\end{figure}

Figure 5(a) shows an example of the Peach-Koehler force, which is similar to an example from previous theoretical work.\cite{Rey1990}  In this example, we have a slab of liquid crystal, with strong surface anchoring that requires $\theta=\pi/4$ at $y=+d/2$, and $\theta=-\pi/4$ at $y=-d/2$.  From one surface to the other, the director field may rotate in two different ways.  In the simplest model with equal Frank constants, these two directions of rotation have the same free energy.  Hence, the system may have coexisting domains with different directions of rotation.  At the interface between the domains, there is a $+1/2$ wedge disclination line in the middle of the cell, with tangent vector $\hat{\mathbf{t}}=\hat{\pmb{\Omega}}=\hat{\mathbf{z}}$.

Now suppose we change the anchoring conditions to $\theta=(\pi/4)+(\delta\theta/2)$ at $y=+d/2$, and $\theta=-(\pi/4)-(\delta\theta/2)$ at $y=-d/2$.  This change requires a torque of $\pmb{\tau}=+(L_x L_z K\delta\theta/d)\hat{\mathbf{z}}$ at $y=+d/2$, and the opposite torque $\pmb{\tau}=-(L_x L_z K\delta\theta/d)\hat{\mathbf{z}}$ at $y=-d/2$.  This torque can be expressed as an effective stress tensor with $\sigma^\mathrm{eff}_{zy}=K\delta\theta/d$, and all other components zero.  Hence, Eq.~(\ref{nematicPK}) gives a Peach-Koehler force (per length) of $\mathbf{f}_\mathrm{PK}=(\pi K\delta\theta/d)\hat{\mathbf{x}}$, which causes the disclination line to move in the $+x$-direction.

This result for the Peach-Koehler force on the disclination line is quite reasonable, because the change of surface anchoring reduces the free energy of the domain on the left, and increases the free energy of the domain on the right.  Hence, the disclination moves to the right in order to expand the lower-energy domain and contract the higher-energy domain.  In previous papers,\cite{Tang2019,Tang2020} we have analyzed this effect as an elastic force, without even mentioning the Peach-Koehler force.  However, it naturally fits into the concept of a Peach-Koehler force.

At this point, the Peach-Koehler force can be generalized to any nematic liquid crystal, even if the director field is out of the $(x,y)$ plane and it depends on all three coordinates $(x,y,z)$.  The rotation vector $\hat{\pmb{\Omega}}$ is still defined, as in the previous sections of this paper, and hence the effective Burgers vector is still $\mathbf{b}^\mathrm{eff}=\pi\hat{\pmb{\Omega}}$.  The effective strain tensor can be generalized to
\begin{equation}
\epsilon^\mathrm{eff}_{ji}=(\hat{\mathbf{n}}\times\partial_i \hat{\mathbf{n}})_j
=\epsilon_{jkl}n_k \partial_i n_l,
\end{equation}
which reduces to Eq.~(\ref{effectivestrain}) if $\hat{\mathbf{n}}=(\cos\theta,\sin\theta,0)$.  (Incidentally, this expression for $\epsilon^\mathrm{eff}_{ji}$ is equivalent to the handedness tensor of Efrati and Irvine.\cite{Efrati2014})  In the simplest model with equal Frank constants, the elastic free energy is
\begin{equation}
F=\frac{1}{2}K(\partial_i n_j)(\partial_i n_j)
=\frac{1}{2}K\epsilon^\mathrm{eff}_{ji}\epsilon^\mathrm{eff}_{ji},
\end{equation}
and hence the effective stress is
\begin{equation}
\sigma^\mathrm{eff}_{ji}=\frac{\partial F}{\partial\epsilon^\mathrm{eff}_{ji}}
=K\epsilon^\mathrm{eff}_{ji}
=K(\hat{\mathbf{n}}\times\partial_i \hat{\mathbf{n}})_j
=K\epsilon_{jkl}n_k \partial_i n_l .
\label{effectivestresswithn}
\end{equation}
This effective stress still has the physical interpretation of a difference of torques across a cell, normalized by the area.  Hence, the nematic Peach-Koehler force (per length) on a local segment of a disclination line can still be constructed as in Eq.~(\ref{nematicPK}).

To demonstrate this generalization, Fig.~5(b) shows an example that is similar to Fig.~5(a) but involves twist.  Here, we have a slab of liquid crystal with strong surface anchoring that requires $\theta=\pi/2$ (so $\hat{\mathbf{n}}=\hat{\mathbf{y}}$) at $z=+d/2$, and $\theta=0$ (so $\hat{\mathbf{n}}=\hat{\mathbf{x}}$) at $z=-d/2$.  In this case, the director may have twist in either direction, and hence the system may have coexisting domains with opposite twist.  At the interface between domains, there is a twist disclination line in the middle of the cell, with tangent vector $\hat{\mathbf{t}}=-\hat{\mathbf{y}}$ and rotation vector $\hat{\pmb{\Omega}}=\hat{\mathbf{z}}$.  Now we twist the top and bottom surfaces in opposite directions, so that $\theta=(\pi/2)+(\delta\theta/2)$ on the top, and $\theta=-(\delta\theta/2)$ on the bottom.  This twist requires a torque of $\pmb{\tau}=+(L_x L_y K\delta\theta/d)\hat{\mathbf{z}}$ at $z=+d/2$, and the opposite torque $\pmb{\tau}=-(L_x L_y K\delta\theta/d)\hat{\mathbf{z}}$ at $z=-d/2$.  This torque can be expressed as an effective stress tensor with $\sigma^\mathrm{eff}_{zz}=K\delta\theta/d$, and all other components zero.  Hence, Eq.~(\ref{nematicPK}) gives a Peach-Koehler force (per length) of $\mathbf{f}_\mathrm{PK}=(\pi K\delta\theta/d)\hat{\mathbf{x}}$, causing the disclination line to move in the $+x$-direction.  Once again, this force can be understood as a result of the reduced free energy for the domain on the left, and the increased free energy for the domain on the right.

For a further example of how the Peach-Koehler force might occur in an experiment, Fig.~5(c) shows a full disclination loop.  In the terminology of Duclos \emph{et al.},\cite{Duclos2020} it is a pure twist loop, and it separates domains of opposite twist inside and outside the loop.  Initially, the top and bottom surfaces have anchoring in perpendicular directions, and hence the opposite domains have equal energy per area.  However, the system is not stable, because the disclination has a line energy $E_\mathrm{line}$ per length.  Because the loop is curved with a radius $R$, the line energy generates a curvature force per length of $f_\mathrm{curv}=E_\mathrm{line}/R$, which points inward.  This force causes the loop to shrink.  However, if we apply a torque to the top and bottom surfaces, the top rotates by $+(\delta\theta/2)$ and the bottom by $-(\delta\theta/2)$.  As a result, the domain inside the loop has a lower energy per area than the domain outside the loop.  Hence, the torque generates a Peach-Koehler force of $f_\mathrm{PK}=(\pi K\delta\theta/d)$, which points outward.  At the radius of $R=(E_\mathrm{line}d)/(\pi K\delta\theta)$, the Peach-Koehler force cancels the curvature force.  This cancellation creates an unstable equilibrium:  If the radius is smaller then the loop shrinks, and if the radius is larger then the loop grows.

\section{Interaction between disclinations}

In the theory of crystalline solids, one important application of the Peach-Koehler force to find the interaction between two dislocations.  Researchers calculate the stress field due to dislocation~1, evaluate it at the position of dislocation~2, and then determine the force acting on dislocation~2.  In this section, we perform an analogous calculation for the interaction between two disclinations in a nematic liquid crystal.  This type of calculation was done by Kl\'{e}man\cite{Kleman1983} for parallel disclinations, but not for non-parallel disclinations.

\begin{figure}
\includegraphics[width=\columnwidth]{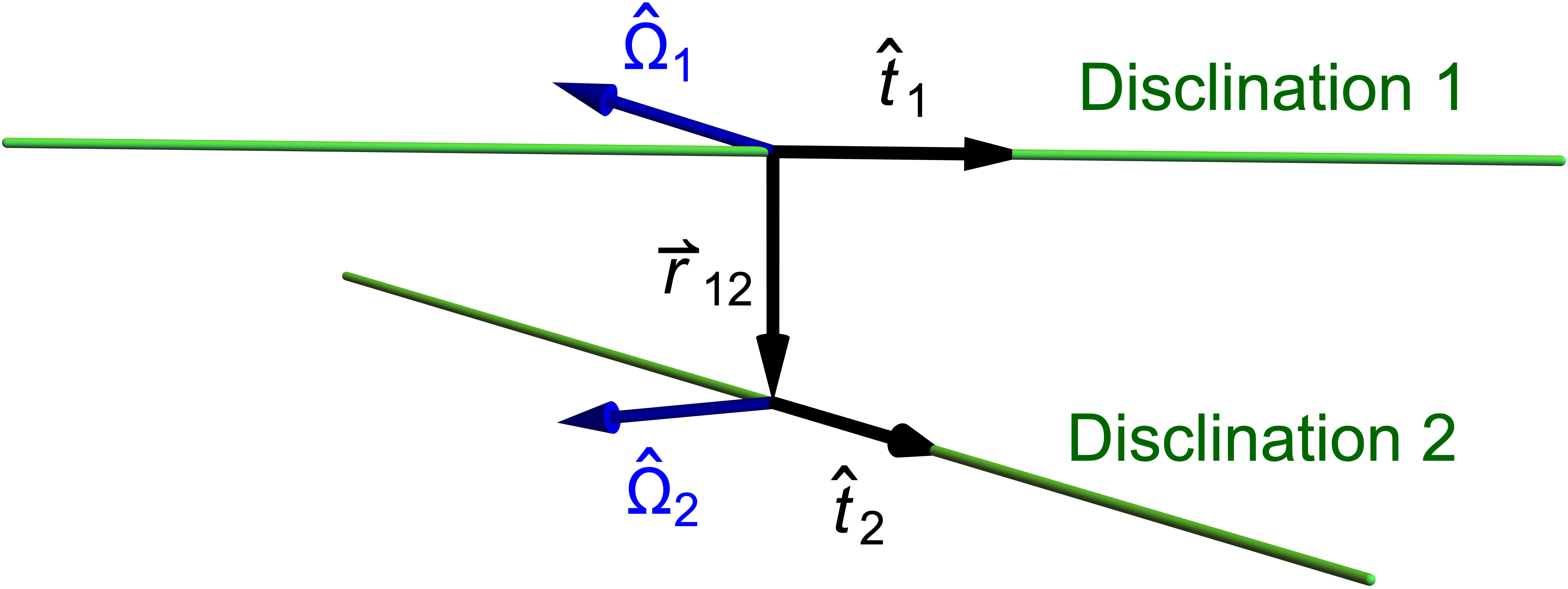}
\caption{Geometry of two interacting disclination lines in a 3D nematic liquid crystal.  The vector $\mathbf{r}_{12}$ goes between the positions of closest approach.}
\end{figure}

For this calculation, we assume that the elastic free energy has the simplest form with equal Frank constants, and that each disclination is a straight line with constant rotation vector and constant tangent vector.  We label the rotation vectors as $\hat{\pmb{\Omega}}_1$ and $\hat{\pmb{\Omega}}_2$, and the tangent vectors as $\hat{\mathbf{t}}_1$ and $\hat{\mathbf{t}}_2$, for disclinations 1 and 2, respectively.  This geometry is shown in Fig.~6.  We choose the origin to be the point on disclination~1 that is closest to disclination~2, and choose the $z$-axis to be aligned with tangent vector $\hat{\mathbf{t}}_1$.  The director field around disclination~1 is then given by Eq.~(\ref{director}).  By differentiating that director field, we obtain
\begin{equation}
\hat{\mathbf{n}}\times\partial_i \hat{\mathbf{n}}=\frac{(-y,x,0)_i}{2(x^2 +y^2)}\hat{\pmb{\Omega}}_1
=\frac{(\hat{\mathbf{z}}\times\mathbf{r})_i}{2[|\mathbf{r}|^2 -(\hat{\mathbf{z}}\cdot\mathbf{r})^2]}\hat{\pmb{\Omega}}_1,
\end{equation}
where $\mathbf{r}=(x,y,z)$.  From Eq.~(\ref{effectivestresswithn}), we can calculate the effective stress tensor induced by disclination~1,
\begin{equation}
(\pmb{\sigma}^\mathrm{eff}_1)_{ji}=K(\hat{\mathbf{n}}\times\partial_i \hat{\mathbf{n}})_j
=\frac{K}{2}(\hat{\pmb{\Omega}}_1)_j \frac{(\hat{\mathbf{z}}\times\mathbf{r})_i}{|\mathbf{r}|^2 -(\hat{\mathbf{z}}\cdot\mathbf{r})^2}
\end{equation}
In general, if disclination~1 has arbitrary tangent vector $\hat{\mathbf{t}}_1$, then the expression becomes
\begin{equation}
(\pmb{\sigma}^\mathrm{eff}_1)_{ji}
=\frac{K}{2}(\hat{\pmb{\Omega}}_1)_j \frac{(\hat{\mathbf{t}}_1 \times\mathbf{r})_i}{|\mathbf{r}|^2 -(\hat{\mathbf{t}}_1\cdot\mathbf{r})^2}
\end{equation}

Now suppose that a small segment of disclination~2 is located at position $\mathbf{r}$, with rotation vector $\hat{\pmb{\Omega}}_2$ and tangent vector $\hat{\mathbf{t}}_2$.  From Eq.~(\ref{nematicPK}), the Peach-Koehler force (per length) acting on this segment is
\begin{equation}
\mathbf{f}_\mathrm{PK}=(\pi\hat{\pmb{\Omega}}_2\cdot\pmb{\sigma}^\mathrm{eff}_1)\times\hat{\mathbf{t}}_2
=\frac{\pi K}{2}(\hat{\pmb{\Omega}}_1 \cdot\hat{\pmb{\Omega}}_2 )
\frac{(\hat{\mathbf{t}}_1\times\mathbf{r})\times\hat{\mathbf{t}}_2}{|\mathbf{r}|^2 -(\hat{\mathbf{t}}_1\cdot\mathbf{r})^2}
\end{equation}
To calculate the total force acting on disclination~2, we must integrate this expression over the entire length of the disclination.  For that integral, we parameterize the position of any segment of disclination~2 as $\mathbf{r}(s)=\mathbf{r}_{12}+s\hat{\mathbf{t}}_2$, where $s$ is the arc length along the disclination, and $\mathbf{r}_{12}$ is the position on disclination~2 that is closest to disclination~1.  Note that $\mathbf{r}_{12}$ is perpendicular to both $\hat{\mathbf{t}}_1$ and $\hat{\mathbf{t}}_2$.  The total force of disclination~1 acting on disclination~2 then becomes
\begin{equation}
\mathbf{F}_{12}=\int_{-\infty}^{\infty}ds\,\mathbf{f}_\mathrm{PK}(s)
=\frac{\pi K}{2}(\hat{\pmb{\Omega}}_1\cdot\hat{\pmb{\Omega}}_2 )
\int_{-\infty}^{\infty}ds
\frac{\mathbf{r}_{12}(\hat{\mathbf{t}}_1 \cdot\hat{\mathbf{t}}_2 )}{|\mathbf{r}_{12}|^2
+s^2[1-(\hat{\mathbf{t}}_1 \cdot\hat{\mathbf{t}}_2 )^2]}.
\end{equation}
If the two disclinations are neither parallel nor antiparallel, so that $(\hat{\mathbf{t}}_1 \cdot\hat{\mathbf{t}}_2 )^2<1$, then this integral can be evaluated as
\begin{equation}
\mathbf{F}_{12}=\frac{\pi^2 K}{2}
\frac{(\hat{\pmb{\Omega}}_1\cdot\hat{\pmb{\Omega}}_2 )(\hat{\mathbf{t}}_1 \cdot\hat{\mathbf{t}}_2 )}{[1-
(\hat{\mathbf{t}}_1 \cdot\hat{\mathbf{t}}_2 )^2 ]^{1/2}}
\frac{\mathbf{r}_{12}}{|\mathbf{r}_{12}|}.
\label{interactionforce1}
\end{equation}
By comparison, if the disclinations are approximately parallel or antiparallel, so that $(\hat{\mathbf{t}}_1 \cdot\hat{\mathbf{t}}_2 )^2\approx1$, then the integral is cut off by the system size $L_\textrm{max}$, and the limits of integration should really be $-L_\textrm{max}/2$ to $L_\textrm{max}/2$.  In that case, the total force is
\begin{equation}
\mathbf{F}_{12}=\frac{\pi K}{2}
(\hat{\pmb{\Omega}}_1\cdot\hat{\pmb{\Omega}}_2 )
(\hat{\mathbf{t}}_1 \cdot\hat{\mathbf{t}}_2 )
\frac{\mathbf{r}_{12}L_\mathrm{max}}{|\mathbf{r}_{12}|^2}.
\label{interactionforce2}
\end{equation}
The crossover between those two expressions occurs at
\begin{equation}
[1-(\hat{\mathbf{t}}_1 \cdot\hat{\mathbf{t}}_2 )^2 ]^{1/2}=\frac{|\mathbf{r}_{12}|}{\pi L_\mathrm{max}}.
\end{equation}

These results for the interaction force have the symmetries that we expect.  The force is invariant if we simultaneously change the signs of $\hat{\pmb{\Omega}}_1$ and $\hat{\mathbf{t}}_1$, or if we simultaneously change the signs of $\hat{\pmb{\Omega}}_2$ and $\hat{\mathbf{t}}_2$.  Furthermore, we can reverse the argument to calculate the force of disclination~2 acting on disclination~1.  The forces are equal and opposite, $\mathbf{F}_{21}=-\mathbf{F}_{12}$, because $\mathbf{r}_{21}=-\mathbf{r}_{12}$.

Because the interaction force is directed along the separation vector $\mathbf{r}_{12}$, it is either repulsive or attractive, depending on the signs of the dot products.  It is repulsive if $(\hat{\pmb{\Omega}}_1\cdot\hat{\pmb{\Omega}}_2 )(\hat{\mathbf{t}}_1 \cdot\hat{\mathbf{t}}_2 )>0$, and attractive if $(\hat{\pmb{\Omega}}_1\cdot\hat{\pmb{\Omega}}_2 )(\hat{\mathbf{t}}_1 \cdot\hat{\mathbf{t}}_2 )<0$.  For non-parallel disclinations, Eq.~(\ref{interactionforce1}) shows that the magnitude of the force is independent of the separation between the disclinations, and it is also independent of the system size.  For parallel or antiparallel disclinations, Eq.~(\ref{interactionforce2}) shows that the magnitude of the force scales as $1/r_{12}$, and it is proportional to the system size, i.e.\ the length of the disclinations.

As a check, we can find the limiting case of the interaction force for an effectively 2D liquid crystal, in which the director field is always in the $(x,y)$ plane and depends only on $x$ and $y$, independent of $z$.  In this system, all disclination lines must be parallel to the $z$-axis, so we can choose the tangent vector $\hat{\mathbf{t}}=\hat{\mathbf{z}}$.  We must calculate the force using Eq.~(\ref{interactionforce2}), and the relevant system size is in the $z$-direction, $L_\mathrm{max}=L_z$.  The only possible types of disclinations are $+1/2$, which have rotation vector $\hat{\pmb{\Omega}}=\hat{\mathbf{z}}$, and $-1/2$, which have $-\hat{\pmb{\Omega}}=\hat{\mathbf{z}}$.  Hence, the interaction force has magnitude $F_{12}=(\pi K L_z)/(2r_{12})$, and it is repulsive for two $+1/2$ or two $-1/2$ disclinations, or attractive for a $+1/2$ and a $-1/2$ disclination.  This result is consistent with well-known theory of 2D liquid crystals.

It might seem surprising that the interaction force depends on the rotation vector $\hat{\pmb{\Omega}}$ but not on the other orientational properties discussed in Secs.~1-3:  the vector $\mathbf{p}$ and the tensors $D_{ij}$ and $T_{ijk}$.  In our previous study of 2D defect orientation,\cite{Tang2017} we found that the well-known interaction force of $F_{12}=(\pi K L_z)/(2r_{12})$ applies only if the two defects have the optimal relative orientation.  If they have a non-optimal relative orientation, then there is an extra term in the interaction, which depends on other orientational properties.  This extra term tends to push the defects into the optimal relative orientation.  Apparently the Peach-Koehler formalism includes an implicit assumption that the disclinations have the optimal relative orientation.  Indeed, a related issue occurs in the theory of crystalline solids.  A dislocation has a Burgers vector, which is a topological invariant, but it also has an extra phase variable, and this phase variable is not included in the Peach-Koehler theory.  Adding the extra phase variable (for crystalline solids) or extra orientational properties (for 3D liquid crystals) remains a further theoretical challenge. 

\section{Active force}

As discussed in the Introduction, much recent research has focused on \emph{active} nematic liquid crystals.  Active liquid crystals are similar in some ways to conventional, passive liquid crystals, but they are not in thermal equilibrium.  Rather, they continually consume energy and convert it into motion.  Many recent studies have shown that activity creates a force proportional to spatial gradients of the nematic director field.  This active force induces the nucleation and motion of disclinations.  Indeed, the motion of disclinations is one of the most prominent features of active liquid crystals.  Hence, it is important to understand how the active motion of disclinations is related to the geometric properties discussed in this article.

The active motion of disclinations was already modeled in the recent article by Binysh \emph{et al}.\cite{Binysh2020}  They begin with the director configuration around a disclination, calculate the active force, put it into the Stokes equations, and solve for the self-propelled velocity of the disclination.  In this section, we repeat their calculation, using a somewhat different method, in order to comment on it.  We obtain essentially the same result as their article, but expressed with the geometric concepts of Secs.~2-4.

In an active liquid crystal, the active force density (per volume) acting on the flow velocity field can be expressed as
\begin{equation}
\mathbf{f}_\mathrm{active}(\mathbf{r})=-\zeta\pmb{\nabla}\cdot(\hat{\mathbf{n}}\hat{\mathbf{n}})
=\zeta[\mathbf{B}(\mathbf{r})-S(\mathbf{r})\hat{\mathbf{n}}(\mathbf{r})].
\end{equation}
Here, $\zeta$ is the activity coefficient, with $\zeta>0$ representing extensile activity, and $\zeta<0$ representing contractile activity.  The modes $\mathbf{B}(\mathbf{r})$ and $S(\mathbf{r})\hat{\mathbf{n}}(\mathbf{r})$ are the bend and splay deformations discussed in Sec.~2.  Note that they are the local, position-dependent deformations, not the averages $\overline{\mathbf{B}}$ and $\overline{S\hat{\mathbf{n}}}$.  This active force induces a local flow velocity field $\mathbf{u}(\mathbf{r})$.  The Stokes equation for the flow velocity field then becomes
\begin{equation}
\mathbf{f}_\mathrm{active}(\mathbf{r})+\mu\nabla^2 \mathbf{u}(\mathbf{r})-\pmb{\nabla}p(\mathbf{r})=0,
\end{equation}
where $p(\mathbf{r})$ is the pressure field, which enforces the constraint of incompressibility
\begin{equation}
\pmb{\nabla}\cdot\mathbf{u}(\mathbf{r})=0.
\end{equation}
For one solution method, we can perform a Fourier transformation from position $\mathbf{r}$ to wavevector $\mathbf{q}$, and obtain
\begin{equation}
\mathbf{f}_\mathrm{active}(\mathbf{q})-\mu q^2 \mathbf{u}(\mathbf{q})-i\mathbf{q}p(\mathbf{q})=0,
\qquad i\mathbf{q}\cdot\mathbf{u}(\mathbf{q})=0.
\end{equation}
The solutions for velocity and pressure in Fourier space are
\begin{equation}
\mathbf{u}(\mathbf{q})=\frac{1}{\mu q^2}\left[\mathbf{I}-\frac{\mathbf{q}\mathbf{q}}{q^2}\right]\cdot\mathbf{f}_\mathrm{active}(\mathbf{q}),
\qquad p(\mathbf{q})=-\frac{i\mathbf{q}}{q^2}\cdot\mathbf{f}_\mathrm{active}(\mathbf{q}).
\label{Fouriersolution}
\end{equation}
Next, we must perform an inverse Fourier transformation to calculate the velocity field $\mathbf{u}(\mathbf{r})$ in position space.  In particular, we would like to find the velocity field evaluated along the disclination line itself, because that gives the self-propelled velocity $\mathbf{u}_\mathrm{SP}=\mathbf{u}(\mathbf{r}=0)$ of the disclination.

We apply this solution method to the director field of Eqs.~(\ref{director}) and~(\ref{Qoutside}).  For this calculation, we must distinguish carefully between vectors in 3D and vectors in the 2D plane normal to the tangent vector $\hat{\mathbf{t}}$.  Hence, we adopt the convention that Latin indices are range over all three directions, while Greek indices range only over the two directions $x$ and $y$.  We also define $\rho=(r_\mu r_\mu)^{1/2}$ as the radius outward from the disclination line in cylindrical coordinates, and $\delta^\perp_{ij}=\delta_{ij}-t_i t_j$ as the 2D Kronecker symbol.  The director field can then be written compactly in tensor notation as 
\begin{equation}
n_i n_j =\frac{\delta_{ij}-\Omega_i \Omega_j}{2}
+\frac{[v_\mu (m_i m_j -m'_i m'_j)+v'_\mu (m_i m'_j +m'_i m_j)]}{2}\frac{r_\mu}{\rho}.
\end{equation}
In position space, the active force is
\begin{align}
&\mathbf{f}_\mathrm{active}(\mathbf{r})_j=\partial_i(n_i n_j)\\
&=-\frac{\zeta[v_\mu (m_\nu m_j -m'_\nu m'_j)+v'_\mu (m_\nu m'_j +m'_\nu m_j)]}{2}
\left[\frac{\delta^\perp_{\mu\nu}}{\rho}-\frac{r_\mu r_\nu}{\rho^3}\right].\nonumber
\end{align}
In Fourier space, this active force becomes
\begin{align}
&\mathbf{f}_\mathrm{active}(\mathbf{q})_j\\
&=-\frac{\zeta[v_\mu (m_\nu m_j -m'_\nu m'_j)+v'_\mu (m_\nu m'_j +m'_\nu m_j)]}{2}
\frac{(2\pi)^2 q_\mu q_\nu \delta(q_z)}{q_\perp^3},\nonumber
\end{align}
with $q_\perp =(q_\mu q_\mu)^{1/2}$.  Hence, the velocity field in Fourier space is
\begin{align}
\mathbf{u}(\mathbf{q})_i=-&\frac{\zeta[v_\mu (m_\nu m_j -m'_\nu m'_j)+v'_\mu (m_\nu m'_j +m'_\nu m_j)]}{2\mu}\times\nonumber\\
&\times\left[\delta_{ij}-\frac{q_i q_j}{q_\perp^2}\right]
\frac{(2\pi)^2 q_\mu q_\nu \delta(q_z)}{q_\perp^5}
\end{align}
Now we perform the inverse Fourier transformation back into real space.  The integrals diverge with small $q_\perp$, corresponding to large system size.  Hence, we impose an infrared cutoff of $q_\mathrm{min}=1/R_\perp$ on the integrals, and obtain the self-propelled velocity
\begin{align}
(\mathbf{u}_\mathrm{SP})_i ={}&\mathbf{u}(\mathbf{r}=0)_i \\
=&-\frac{\zeta R_\perp[v_\mu (m_\nu m_j -m'_\nu m'_j)+v'_\mu (m_\nu m'_j +m'_\nu m_j)]}{16\mu}\times\nonumber\\
&\times\left[4\delta_{ij}\delta^\perp_{\mu\nu}
-(\delta^\perp_{ij}\delta^\perp_{\mu\nu}+\delta^\perp_{i\mu}\delta^\perp_{j\nu}+\delta^\perp_{i\nu}\delta^\perp_{j\mu})\right].\nonumber
\end{align}
That expression simplifies to
\begin{equation}
\mathbf{u}_\mathrm{SP}=-\frac{\zeta R_\perp}{16\mu}\left[
4\mathbf{p}+4\hat{\mathbf{t}}(\hat{\mathbf{t}}\cdot\mathbf{p})-\hat{\mathbf{v}}\sin^2 \beta\right].
\end{equation}
Recall that $\mathbf{p}$ is the defect orientation vector defined in Eq.~(\ref{pdefinition}), and $\hat{\mathbf{v}}=(\cos\phi_0,\sin\phi_0,0)$ is the direction outward from the disclination such that $\hat{\mathbf{n}}$ is in the plane perpendicular to $\hat{\mathbf{t}}$, as defined in Eq.~(\ref{Qoutside}).  This result for $\mathbf{u}_\mathrm{SP}$ is equivalent to the expression found previously by Binysh \emph{et al}.\cite{Binysh2020}

To interpret this result, note that the incompressibility constraint plays an important role in the calculation.  In Fourier space, Eq.~(\ref{Fouriersolution}) shows that $\mathbf{f}_\mathrm{active}=\zeta{[\mathbf{B}-S\hat{\mathbf{n}}]}$ provides the force, but the direction of the velocity is not the same as the direction of the force.  Rather, the incompressibility constraint gives the projection operator $\mathbf{I}-\mathbf{q}\mathbf{q}/q^2$, which acts on the force to give the velocity.  The same effect can be seen in real space.  If we did not have the incompressibility constraint, and hence did not have the projection operator, then the self-propelled velocity would be simply
\begin{equation}
\mathbf{u}_\mathrm{unconstrained}=-\frac{\zeta R_\perp}{2\mu}\mathbf{p}.
\end{equation}
Hence, we can say that the active force acting on the disclination is in the $-\zeta\mathbf{p}$ direction, which is consistent with the average bend and average splay of Sec.~2.  However, the incompressibility constraint prevents the disclination from moving in that direction, because such motion would induce density changes.  The difference between $\mathbf{u}_\mathrm{SP}$ and $\mathbf{u}_\mathrm{unconstrained}$ is a vector perpendicular to $\hat{\mathbf{t}}$.  This result is reasonable, because a force parallel to $\hat{\mathbf{t}}$ does not induce any density changes in this model of a uniform, straight disclination line.

In conclusion, this article has analyzed the geometry of disclination lines in 3D nematic liquid crystals, as well as the forces acting on these defects.  Our geometric analysis shows that disclination lines have features with one-fold, two-fold, and three-fold symmetry.  The one-fold and three-fold symmetric features correspond to orientational properties of $+1/2$ and $-1/2$ disclinations in 2D nematic liquid crystals, while the two-fold features are a new aspect of twisted disclinations in 3D.  Using this geometric analysis, we have investigated three types of forces:  Peach-Koehler forces due to externally applied stress, interaction forces between two disclination lines, and active forces.  We find that Peach-Koeher and interaction forces are sensitive to the disclination rotation vector $\hat{\pmb{\Omega}}$, while active forces depend on higher-order geometric features, particularly the vector $\mathbf{p}$, which corresponds to the orientation of a $+1/2$ disclination in 2D.  This characterization of forces should facilitate the analysis of experiments on both conventional and active liquid crystals.

\section*{Conflicts of interest}

There are no conflicts to declare.

\section*{Acknowledgements}

We would like to thank S~Afghah for a performing a previous version of the simulations in Sec.~4.  This work was supported by National Science Foundation Grant No.~DMR-1409658. 





\bibliography{disclinationlines1} 
\bibliographystyle{rsc} 

\end{document}